\documentclass[
  aps,
  pre,
  longbibliography,
  reprint,
  superscriptaddress,
  amsmath,
  amssymb,
  floatfix
]{revtex4-2}
\usepackage{graphicx}
\usepackage{dcolumn}
\usepackage{bm}
\usepackage{bbm}
\usepackage{array}
\usepackage{tabularx}
\usepackage{booktabs}
\usepackage{xcolor}
\usepackage{url}
\usepackage{enumitem}
\usepackage[caption=false,font=footnotesize]{subfig}
\usepackage{hyperref}
\hypersetup{hidelinks}

\allowdisplaybreaks

\begin{document}

\title{Overload-Based Cascades in Multiplex Flow Networks with Partial Functionality}

\author{Orkun \surname{\.{I}rsoy}}
\email{oirsoy@andrew.cmu.edu}
\affiliation{
Department of Electrical and Computer Engineering,
Carnegie Mellon University,
Pittsburgh, Pennsylvania 15213, USA
}

\author{Osman \surname{Ya\u{g}an}}
\email{oyagan@andrew.cmu.edu}
\affiliation{
Department of Electrical and Computer Engineering,
Carnegie Mellon University,
Pittsburgh, Pennsylvania 15213, USA
}

\date{\today}

\begin{abstract}
Cascading failures driven by load/flow redistribution are widespread in networked systems such as power grids, supply chains, and cloud computing centers. 
Most existing flow-network models assume that a node either functions or fails as a whole (which we refer to as \textit{joint functionality}), but in many real systems a node supports several distinct flows/functionalities that share node-level resources, and failure in one of them does not necessarily imply failure in the others. 
We study this setting through multiplex flow networks with \textit{partial functionality}, in which nodes' functionalities share limited resources, but a node can remain operational in some functionalities while having failed in others.
Due to shared resources, a heavy load on one functionality reduces the capacity available to the others, which is quantified by cross-layer influence factors.
When a node fails in one layer, its load is redistributed among the surviving nodes in that layer, while the node may continue to operate in the others. 
Using mean-field analysis, we derive recursive equations for the final system sizes, i.e. the fraction of surviving nodes in each layer after the cascade stops, as a function of the initial fraction of failed nodes and the joint distribution of initial loads and capacities. 
We validate our analysis across several initial load–capacity distributions through simulations, and then analyze several characteristics of the cascade dynamics, such as non-monotone robustness curves, different cascade-outcome regimes, and their relation with increased cross-layer influence. 
We map the cascade outcomes to distinct steady-state regimes, including single-layer survival phases that are absent in joint-functionality models, and we show that partial functionality can increase robustness relative to the joint-functionality case. 
Finally, we study robustness maximization under a fixed total capacity budget by comparing several capacity allocation strategies.
We propose a strategy that combines the cross-layer influence with local neighborhood information on load and degree, and show that it achieves the strongest robustness performance across the configurations considered.
\end{abstract}


\maketitle

\section{Introduction}

In October 2025, a major outage in Amazon Web Services (AWS) disrupted internet platforms worldwide after a routing fault caused congestion and traffic redistribution, overloading multiple servers and triggering cascading service disruptions~\cite{Hicks_2025}. 
Earlier that year, in April 2025, an overload in the Iberian power grid initiated a cascade of generation losses and load shedding, resulting in a large-scale blackout across Spain and Portugal~\cite{ENTSOE_Iberian_Blackout_2025}. 
In both events, the failure of one component transferred excess load to dependent components, propagating through the network and leading to a system-wide collapse: the Iberian blackout alone affected tens of millions of people and severely interrupted transportation, telecommunications, and industrial activities across the region~\cite{ENTSOE_Iberian_Blackout_2025}.
Such \textit{cascading failures}~\cite{Buldyrev} are a recurring hazard in a diverse range of \textit{networked systems} such as supply chains~\cite{supply_chain_example,supply_chain_ex} and cloud data centers~\cite{cloud_computing_example}, and understanding how they propagate is essential for designing more robust infrastructure.

Existing studies on cascading failures consider various mechanisms of failure propagation. 
The \textit{percolation-based models} \cite{Buldyrev,percolation_1,percolation_2,percolation_3,percolation_4,percolation_5} focus on structural connectedness and apply to networks where system functionality depends on mutual reachability.
These models are particularly suitable for communication and cyber-physical systems, where the loss of connectivity directly impairs operation.  
A separate line of work focuses on \textit{overload-based} or \textit{flow-redistribution} models \cite{flow_1,flow_2,flow_3,flow_4,single_flow_optimizing,Chen_Hu_Meng_Yu_2024,alpha1,alpha2}, which capture systems governed by physical or workload flows, such as power transmission, traffic, or distributed computing. 
In flow networks, failure of a component typically causes redistribution of its load among surviving components, which may result in overload and subsequent failures.  

Most existing studies on the robustness of flow networks have focused on single-layer structures \cite{flow_1,flow_2,flow_3,flow_4,single_flow_optimizing,Chen_Hu_Meng_Yu_2024,alpha1,alpha2} or on interconnected multilayer systems in which flow can be redistributed across layers \cite{scala,zhang_two_flow_redistribution,Kumar_Kumari_Bala_2021,pei_et.al.,hong_suppressing_2016,Wang_Jin_Zhao_2021}. 
These approaches, however, do not capture settings where multiple functionalities coexist on the same node and compete for its finite resources. 
This limitation motivates the study of {\em multiplex flow networks}, in which each node simultaneously supports several distinct flows and its operational state depends on load-capacity conditions that couple the layers through shared resources.
In cloud infrastructures, for example, a server may handle GPU-intensive inference and memory-intensive data services in parallel, such that heavy demand in one task degrades performance in another \cite{cloud_computing_example}. 
Similar patterns arise in management systems where each unit handles multiple tasks, and in supply-chain networks where facilities oversee the production and distribution of multiple products.

Despite the relevance of multiplex flow networks, research in this area remains limited.  
Existing studies either focus on specific coupling strategies between dual layers \cite{Zhou_Elmokashfi_2017,Ma_Xin_2024}, where node dependencies are modeled through topological connections or initial load assignments, or extend classical single-layer models, such as the sandpile model, to multiplex settings where cascades are triggered by random load increments \cite{Lee_Goh_Kim_2012}.  
A recent study \cite{WANG2025130373} proposed an interdependent overload model in which the load in one layer influences the effective capacity in another, highlighting the impact of cross-layer influence on failure propagation.  
However, this exploration remains limited to empirical datasets and does not generalize to arbitrary load–capacity distributions or analytical formulations of the cascade dynamics.  

In our prior work \cite{irsoy2025}, we introduced a multiplex flow network model in which the overload condition of a node depends on its loads in all layers.  
In that formulation, the functionality of a node was fully coupled across layers; that is, if a node failed in one layer, it simultaneously failed in all others, in line with the other network models from the literature \cite{Zhou_Elmokashfi_2017,lee2016strength}. 
This assumption, however, overlooks scenarios where nodes can partially function in one layer while failing in the other, even though functionalities share common resource constraints.
A practical example is 5G network slicing, where distinct virtual networks share the same physical infrastructure but serve different purposes~\cite{5G_network_slice}.
Heavy traffic in one slice can reduce the capacity available to others, yet a slice can remain operational as long as its own requirements are met.
A similar pattern arises in supply chain networks, where a facility handling multiple product lines may suspend operations for one product under resource pressure while continuing to serve others.
In both cases, tasks share common resources, but failure in one layer does not have to shut down the others. We refer to this as \textit{partial functionality}.

In this paper, we introduce \textit{multiplex flow networks with partial functionality}, in which a node's functionalities share limited resources, but a node can remain operational in some functionalities while failing in others.
In this setting, the survival of each node in a given layer depends on the load it supports across all layers, reflecting the effect of shared resource constraints.
To quantify this, we use \textit{cross-layer influence factors} that represent the impact of the load in one layer on the capacity in other layers.
When a node fails in one layer, its load in that layer is  redistributed among the remaining active nodes within the same layer, while it continues to operate in other layers as long as the corresponding survival conditions are not violated.  
This mechanism captures a broad class of systems where functional interdependencies arise from shared node-level resources rather than from direct operational coupling.
The setting is qualitatively different from the \textit{joint-functionality overload condition} ~\cite{irsoy2025} and from related multiplex overload models~\cite{Zhou_Elmokashfi_2017,Lee_Goh_Kim_2012,lee2016strength}.

We analyze overload-based cascading failures initiated by random attacks under the global redistribution rule, where the load of each failed node is distributed evenly among the surviving nodes in the same layer. 
Using mean-field analysis, we derive recursive equations for the final system sizes as functions of the attack size, the joint load and capacity distribution, and the cross-layer influence factors. 
We focus on the two-layer case, with layers-$A$ and -$B$, in which partial functionality leads to distinct single layer and both layer survival states; the recursive equations track each layer separately, distinguishing nodes that function in both layers from those functioning only in layer-$A$ or only in layer-$B$. 
We validate these equations through numerical simulations across a diverse set of initial load and capacity distributions and cross-layer influence factors.

Our analysis shows that partial functionality leads to qualitatively new cascading behaviors including asymmetric layer collapses, where one layer fails while the other keeps functioning, and non-monotone robustness curves, where a layer can reach a higher final surviving fraction under a larger attack. 
We trace these effects to the timing of the cascade and to the capacity release mechanism, i.e., when a node fails in one layer, some capacity becomes available in the other layer through cross-layer influence. 
We then present phase diagrams over the attack size and the cross-layer influence factors, which map distinct cascade-outcome regimes, including both-layer and single-layer survival phases that do not appear in joint-functionality models. 
They also show that cross-layer influence can act {\em asymmetrically} across layers, improving the survival of one layer while harming the other, an emergent effect of the interplay between cross-layer influence and the underlying load and capacity distributions. 
We further show that partial functionality increases robustness relative to the joint-functionality case not only by allowing single-layer survival states, but also by expanding the both-layer survival region.

Finally, we study robustness maximization under a limited total-capacity constraint. 
We evaluate several capacity allocation strategies under both global and local redistribution, and we examine the trade-offs between overall system performance and the robustness of individual layers. 
Motivated by our findings under global redistribution, we propose a capacity allocation strategy that assigns capacity based on first-degree neighbor load, node degree, and cross-layer influence. 
Across all tested configurations and network topologies (including Erd\H{o}s--R\'enyi graphs and scale-free networks), this strategy yields the highest robustness in our experiments.

The remainder of this paper is organized as follows.  
Section~II presents the partial-functionality multiplex flow network framework and defines the partial-functionality overload condition.
Section~III derives the analytical framework and recursive equations governing the cascade dynamics.  
Section~IV reports numerical results, including validation, phase diagrams, and $\beta$ sweeps.
Section~V examines free-space allocation strategies under global and local redistribution.
Section~VI concludes the paper with key insights and future research directions.

\section{System Model}
\label{sec:model}

\subsection{Multiplex Flow Network Model}

\begin{figure}[ht]
    \centering
    \includegraphics[width=0.65\linewidth]{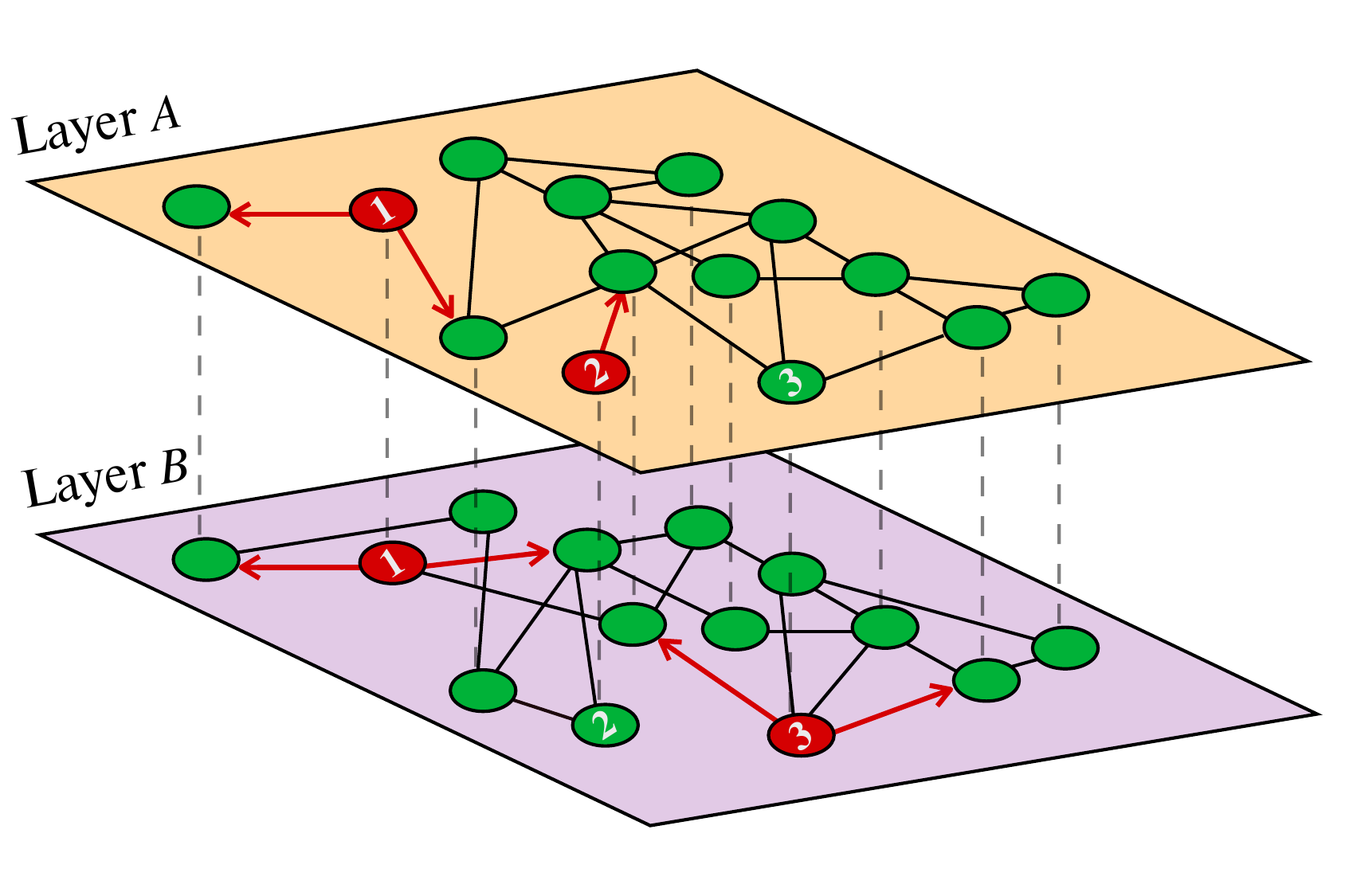}
    \caption{Multiplex flow network with partial functionality. Each layer is defined on the same set of vertices but with (potentially) different edge sets, and each layer is responsible for carrying/supplying a different flow type. The highlighted nodes $v_1$, $v_2$, and $v_3$ carry load $L_{x,A}$ in layer-$A$ and $L_{x,B}$ in layer-$B$ for $x\in\{ 1,2,3\}$. Node $v_1$ exemplifies a node failing in both layers hence both $L_{1,A}$ and $L_{1,B}$ is redistributed in the respective layers. $v_2$ and $v_3$ represent the nodes failing partially, failed in layer-$A$ and layer-$B$ respectively, hence their load in the failed layer is redistributed only in that layer while they continue functioning in the other.}
    \label{fig:multiplex_flow_network_fig}
\end{figure}

Consider a network where a set of nodes $\mathcal{N} = \{1,2,\dots,N\}$ is responsible for transporting (or supporting) $M\geq2$ distinct types of flows (or tasks/functionalities), each supported by a different graph structure. 
This structure is represented as a multiplex network with $M$ layers, where each layer corresponds to a specific flow type.
We focus on the two-layer case, denoted as $\mathcal{M} = \{A,B\}$, demonstrated in Figure~\ref{fig:multiplex_flow_network_fig}, since the same framework can in principle be extended to larger $M$, but the number of possible node states grows as $M$ increases, making the recursive analysis increasingly cumbersome.

\begin{figure}[b]
\centering
\subfloat[Partial-functionality overload condition.]{%
    \includegraphics[width=0.48\linewidth]{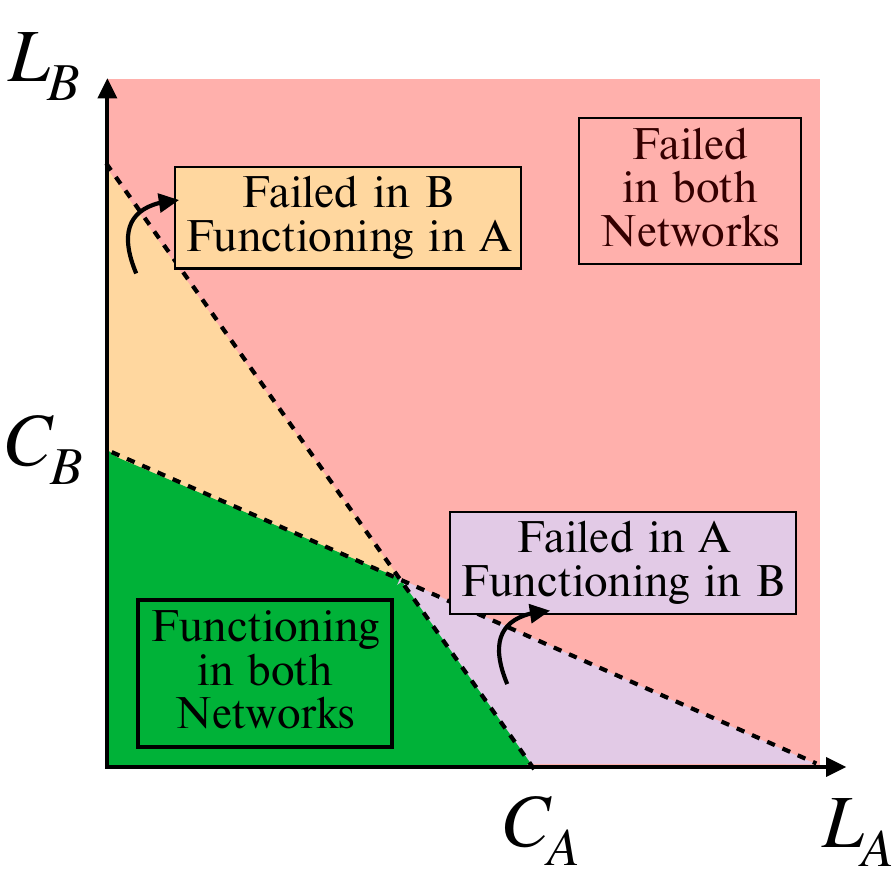}}
\subfloat[Joint-functionality overload condition.]{%
    \includegraphics[width=0.48\linewidth]{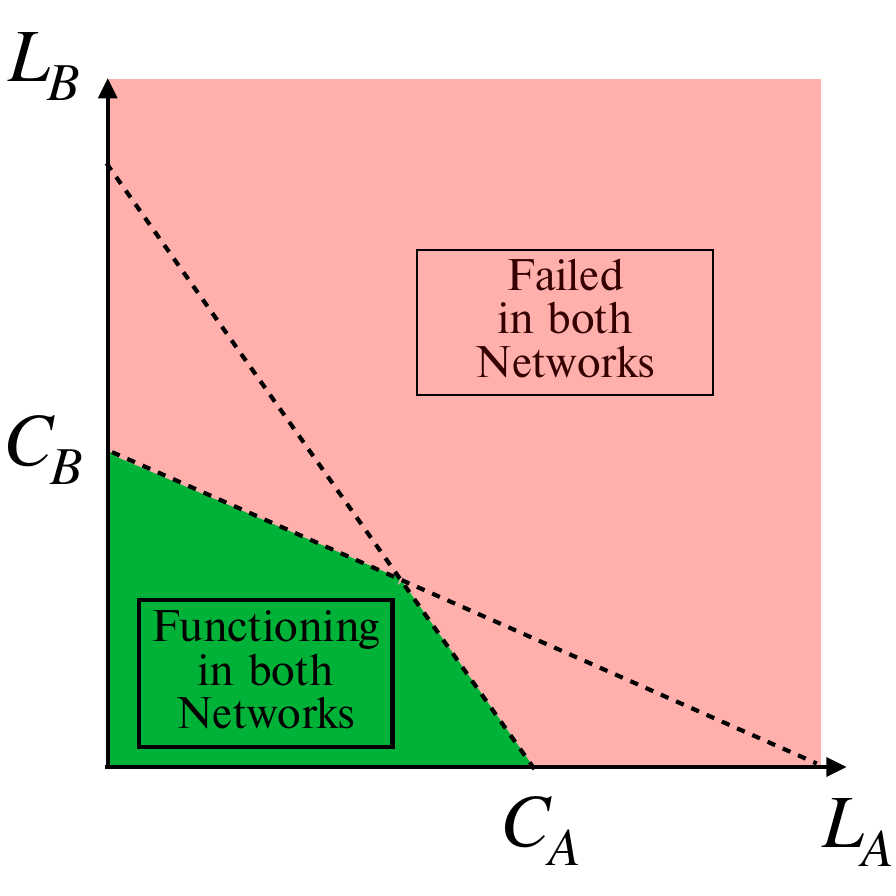}}
\caption{\label{fig:failure_conditions}
Failure/overload conditions defined as a partitioning of the $(L_A,L_B)$ plane.
(a) Partial-functionality overload condition.
(b) Joint-functionality overload condition~\cite{irsoy2025}.
}
\end{figure}

Each node \( v_x \) carries a flow (or load) represented by the vector \( \mathbf{L}_x = [L_{x,A}, L_{x,B}] \).
Under partial functionality, each node falls into one of four states: surviving in both layers, in layer-$A$ only, in layer-$B$ only, or failed in both.
Figure~\ref{fig:failure_conditions}(a) illustrates such a partition for a single node, distinguishing the values of $[L_{x,A}, L_{x,B}]$ that would result in failure in one or both layers.
The capacity vector \( \mathbf{C}_x = [C_{x,A}, C_{x,B}] \) specifies the maximum load the node can sustain in each layer when the load from the other layer is zero.
Failures are determined by the overload conditions introduced below, and failed loads are subsequently redistributed to surviving nodes, possibly causing additional failures.
Since the model assumes that each functionality or flow is distinct, there is no transfer of load between layers; instead, the load redistribution is confined within each layer.  

In this model, survival in each layer is governed by the \textit{partial-functionality overload condition} (Fig.~\ref{fig:failure_conditions}(a)). 
The \textit{cross-layer influence factors} $\beta_A$ and $\beta_B$ represent the unit impact of the load in one layer on the other, and the two survival inequalities are evaluated independently.
A node survives in layer-$A$ if the first inequality holds and in layer-$B$ if the second holds:
\begin{align}
    & L_{x,A} + \beta_B L_{x,B} \leq C_{x,A}, 
    \label{eq_main:failure_condition_new_A}\\
    & L_{x,B} + \beta_A L_{x,A} \leq C_{x,B}
    \label{eq_main:failure_condition_new_B}
\end{align}
Thus, the load in layer-$A$ contributes partially to the effective load in layer-$B$, and vice versa.
For example, if a node $v_x$ satisfies the first but not the second, it fails in layer-$B$ (so $L_{x,B}$ is set to zero and redistributed among the surviving nodes in layer-$B$), but it continues to function in layer-$A$.  
When a node fails in one layer, its load is redistributed to other functioning nodes in that layer, and subsequently is set to zero, so it no longer contributes to overload in the other.

For comparison, Fig.~\ref{fig:failure_conditions}(b) shows the \textit{joint-functionality overload condition} studied in our prior work~\cite{irsoy2025}: the same cross-layer influence exists, but failure in one layer for a node implies failure in all layers.
In that setting, a node survives only if both inequalities, \eqref{eq_main:failure_condition_new_A} and \eqref{eq_main:failure_condition_new_B}, hold; if either is violated, the node fails in all layers and its loads are redistributed within each layer.
Despite easing the analytical tractability, it leads to a monotone effect of \textit{cross-layer influence}: increasing the cross-layer influence shrinks the dual-layer functional region and therefore reduces robustness.
By contrast, \textit{partial functionality} introduces layer-specific survivability regions, creating a trade-off between single-layer and dual-layer functional states rather than a purely monotone effect.
This motivates an analytical characterization of this trade-off, even though it imposes additional complexity in the derivations.

\subsection{Redistribution Rule and Robustness Metrics}
In our analysis, we adopt the \textit{global redistribution rule}, where the load of a failed node is evenly distributed among all surviving nodes.  
Specifically, if a node $v_x$ fails in a specific layer, each type-$i$ flow, $L_{x,i}$, is redistributed within its respective layer-$i$.  
This assumption is widely used in overload-based cascading failure models because it captures \emph{long-range redistribution effects}, where the failure of a single node can influence the entire system through global rebalancing of workload \cite{flow_1,scala,single_flow_optimizing}.  
Such global effects are characteristic of systems in which tasks or demands can be reassigned without strong spatial or topological constraints.  
For example, during the 2025 AWS \cite{Hicks_2025} outage, the failure of a small set of components triggered system-wide rerouting of requests—placing additional stress on distant resources and leading to further disruptions.  
This illustrates how failures in large-scale flow systems often propagate through global workload shifts rather than only through local interactions.
In addition, the global rule yields a tractable analytical formulation: all surviving nodes in a layer receive the same additional load, allowing the cascade to be described through recursions for the excess loads and surviving fractions.  
This preserves the essential behavior of intermittent failures while keeping the model analytically tractable under general load-capacity distributions.

While the global rule is useful for analysis, it does not incorporate network structure, since redistributed load is shared regardless of connectivity.
To represent settings where redistribution is constrained by topology, we also consider a \textit{local redistribution rule}, where the load of a failed node in a layer is redistributed only among its surviving neighbors in that layer.
Our analytical results focus on the global redistribution rule, but we complement them with simulations under local redistribution in Section~\ref{sec:fsa} and use these experiments to evaluate robustness and free-space allocation strategies when load redistribution is localized.

To analyze the system’s robustness, we examine cascading failures initiated by \textit{random attacks}.  
An initial attack removes a fraction $p \in (0,1)$ of nodes, i.e., failing them in both layers, which triggers load redistribution and potential secondary failures.  
In the steady state, let $\mathcal{N}_{A}(p)$ and $\mathcal{N}_{B}(p)$ denote the sets of nodes that survive in layers $A$ and $B$, respectively. 
We further define $\mathcal{N}_{AB}(p):=\mathcal{N}_{A}(p)\cap \mathcal{N}_{B}(p)$ as the set of nodes that survive in both layers.
When needed we use, $\mathcal{N}_{A\setminus B}(p):=\mathcal{N}_{A}(p)\setminus \mathcal{N}_{B}(p)$ and $\mathcal{N}_{B\setminus A}(p):=\mathcal{N}_{B}(p)\setminus \mathcal{N}_{A}(p)$.
We quantify system robustness using the steady-state fraction of surviving nodes in both layers ($n_{AB,\infty}(p)$), in layer-$A$ ($n_{A,\infty}(p)$), and in layer-$B$ ($n_{B,\infty}(p)$):  
\begin{equation}
    n_{i,\infty}(p) := \lim_{N\to\infty} \frac{E \left[|\mathcal{N}_{i}(p)|\right]}{N} \quad \text{for } i\in\{A,B,AB\}
    \label{eq:n_infinity}
\end{equation}

Throughout the paper, we analyze the final system size $n_{i,\infty}(p)$ for $i \in \{A, B, AB\}$ under any attack size $0 < p < 1$.  
Another key objective is to determine the \textit{critical attack size} $p^*$ beyond which a given layer, or both-layer operation, ceases to survive.
We define separate critical attack sizes:
\begin{align}  
    p^*_A &:= \sup\{p : n_{A,\infty}(p) > 0\}, \label{eq:critical_p_a}\\
    p^*_B &:= \sup\{p : n_{B,\infty}(p) > 0\}, \label{eq:critical_p_b}\\
    p^*_{AB} &:= \sup\{p : n_{AB,\infty}(p) > 0\}.
    \label{eq:critical_p_ab}
\end{align}

Here, $p^*_A$ and $p^*_B$ denote the attack sizes beyond which all nodes in layer-$A$ and layer-$B$ fail, respectively.  
In comparison, $p^*_{AB}$ represents the attack size above which no node remains functional in \emph{both} layers at the same time.  
Therefore, it follows that $p^*_{AB} \leq p^*_A$ and $p^*_{AB} \leq p^*_B$.

\subsection{Cascade Dynamics}

We model the cascade in discrete iterations $t=0,1,2,\dots$, where each iteration represents a round of flow redistribution.
At $t=0$, an initial random attack removes a fraction $p$ of nodes uniformly at random, and their loads are redistributed within the corresponding layer.
Subsequently, at each iteration, nodes that violate the overload conditions are removed simultaneously, their failed-layer loads are redistributed within that layer, and the process repeats until no further overload-induced failures occur.

Each node $v_x$ operates in layers $A$ and $B$ and carries an initial load vector $[L^{(0)}_{x,A},L^{(0)}_{x,B}]$.
Under the \textit{global redistribution rule}, all surviving nodes in the same layer receive the same additional burden at iteration $t$.
Denoting the \emph{excess load per surviving node} by $Q_{A,t}$ and $Q_{B,t}$ in layers $A$ and $B$, respectively, the instantaneous loads of a surviving node at iteration $t$ are
\begin{equation}
\begin{array}{c}
     L_{x,A}^{(t)} = L^{(0)}_{x,A} + Q_{A,t},\\
     L_{x,B}^{(t)} = L^{(0)}_{x,B} + Q_{B,t},
\end{array}
\label{eq:inst_loads}
\end{equation}
and $L_{x,i}^{(t)}=0$ once the node has failed in layer $i$.
Given the instantaneous loads, the overload conditions for layers $A$ and $B$ are
\begin{align}
    L_{x,A}^{(t)} + \beta_B L_{x,B}^{(t)} &> C_{x,A}, \label{eq:failure_A}\\
    L_{x,B}^{(t)} + \beta_A L_{x,A}^{(t)} &> C_{x,B}. \label{eq:failure_B}
\end{align}
Thus, nodes satisfying \eqref{eq:failure_A} fail in layer-$A$, and nodes satisfying \eqref{eq:failure_B} fail in layer-$B$.

Rather than specifying capacities directly, we follow the load--free-space formulation used in overload cascade models~\cite{irsoy2025,zhang_two_flow_redistribution,ozel_2018}: for each node $v_x$ we define two nonnegative quantities per layer: its baseline (initial) load and its free space, denoted $L^{(0)}_{x,A},L^{(0)}_{x,B}$ and $S_{x,A},S_{x,B}$ for layers $A$ and $B$, respectively.
The free space $S_{x,i}$ represents the additional load that $v_x$ can tolerate in layer $i$ beyond its baseline load plus the cross-layer contribution.
Capacities are then
\begin{equation}
\begin{aligned}
C_{x,A} &= L^{(0)}_{x,A} + \beta_B L^{(0)}_{x,B} + S_{x,A},\\
C_{x,B} &= L^{(0)}_{x,B} + \beta_A L^{(0)}_{x,A} + S_{x,B}.
\end{aligned}
\label{eq:capacity_ab}
\end{equation}
We assume that $(L^{(0)}_{x,A},S_{x,A},L^{(0)}_{x,B},S_{x,B})$ are i.i.d.\ across nodes with joint CDF $ P_{L_A S_A L_B S_B}(y_{L_A},y_{S_A},y_{L_B},y_{S_B})
= P\!\left[L_A\le y_{L_A},\, S_A\le y_{S_A},\, L_B\le y_{L_B},\, S_B\le y_{S_B}\right]$
 and associated joint PDF $p_{L_A S_A L_B S_B}$.
We further assume positive support, i.e., $L^{(0)}_{x,A}>0$, $S_{x,A}>0$, $L^{(0)}_{x,B}>0$, and $S_{x,B}>0$ for all $v_x$, which ensures all nodes are functioning prior to the attack while allowing arbitrary nonnegative joint distributions to capture load–capacity relationships.

With these definitions, we revisit the overload conditions \eqref{eq:failure_A}–\eqref{eq:failure_B}.
If a node $v_x$ is functioning in both layers, both conditions must be checked to determine its state in the next iteration.
Substituting \eqref{eq:inst_loads} and \eqref{eq:capacity_ab} into \eqref{eq:failure_A}–\eqref{eq:failure_B} cancels the baseline terms and yields the free-space comparisons:
\begin{align}
    S_{x,A} &> Q_{A,t} + \beta_B Q_{B,t}, \label{eq:survival_A}\\
    S_{x,B} &> Q_{B,t} + \beta_A Q_{A,t}. \label{eq:survival_B}
\end{align}
If both are satisfied, the node continues to function in both layers; otherwise, the load in the failed layer(s) is set to zero and redistributed in the respective layer(s).
This motivates the notation of \emph{effective excess loads}
\begin{equation}
\begin{array}{c}
Q'_{A,t} = Q_{A,t} + \beta_B Q_{B,t}, \\
Q'_{B,t} = Q_{B,t} + \beta_A Q_{A,t},
\end{array}
\label{eq:effective_excess_loads}
\end{equation}
under which the survival checks for a node that is alive in both layers reduce to
\begin{align}
S_{x,A} &> Q'_{A,t} \quad \text{(survive in layer-$A$ at $t$)}, \label{eq:survival_short_A}\\
S_{x,B} &> Q'_{B,t} \quad \text{(survive in layer-$B$ at $t$)}. \label{eq:survival_short_B}
\end{align}

If a node $v_x$ is functioning in only one layer, a single condition determines its survival in the respective layer.
Consider the case where node $v_x$ only functions in layer-$A$, so $L^{(t)}_{x,B}=0$.
Then substituting \eqref{eq:inst_loads} and \eqref{eq:capacity_ab} into \eqref{eq:failure_A} yields
\begin{equation}
    S_{x,A} > Q_{A,t} - \beta_B L^{(0)}_{x,B},
    \label{eq:survival_short_single}
\end{equation}
with the symmetric expression for survival in $B$ once $A$ has failed.

\section{Analytical Results} 
\label{sec:anlysis}

In this section, we provide a mean-field analysis of cascading failures under the partial-functionality model introduced in Section~\ref{sec:model}.

\begin{figure}[t!]
    \centering
    \includegraphics[width=\linewidth]{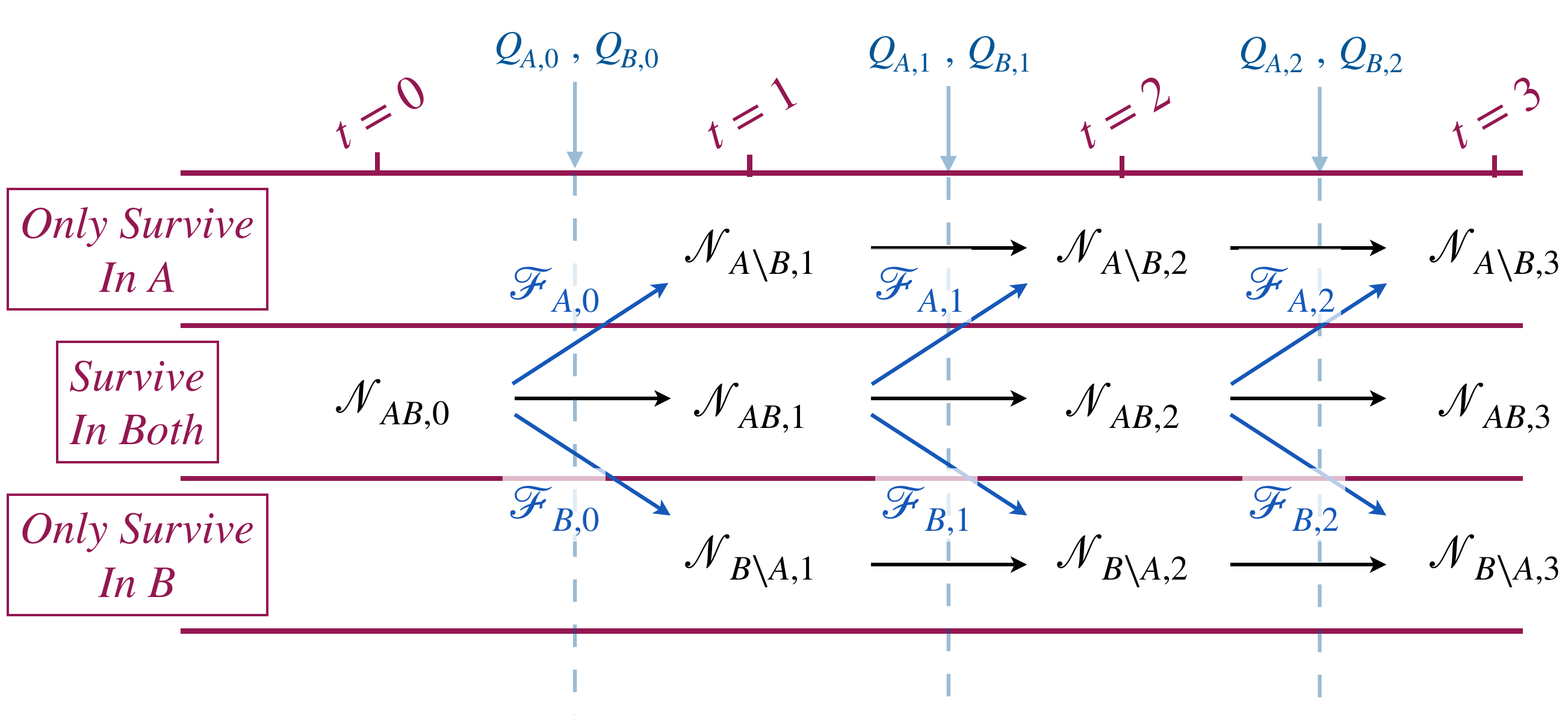}
    \caption{Conceptual diagram of the cascade progression. Here $t=0,1,2,\cdots$ indicates the rounds of flow redistribution. 
$\mathcal{N}_{AB,t}$ denotes the set of nodes surviving in both layers at iteration $t$. 
For $i\in\{A,B\}$, $\mathcal{N}_{i,t}$ denotes the set of nodes surviving in layer $i$ at iteration $t$, and $\mathcal{N}_{i\setminus j,t}:=\mathcal{N}_{i,t}\setminus \mathcal{N}_{j,t}$ denotes the set of nodes surviving only in layer $i$ (with $j\neq i$). 
Moreover, $\mathcal{F}_{i,t}$ is the set of nodes that were surviving in both layers at iteration $t$ but will survive only in layer $i$ at iteration $t{+}1$. 
$Q_{A,t},Q_{B,t}$ denote the excess load per surviving node resulting from the failures up to iteration $t$.}

    \label{fig:flow diagram}
\end{figure}

We track the cascade through the sets $\mathcal{N}_{A,t}$ and $\mathcal{N}_{B,t}$ of nodes operating in layers $A$ and $B$ at iteration $t$, and their intersection $\mathcal{N}_{AB,t}:=\mathcal{N}_{A,t}\cap \mathcal{N}_{B,t}$ as summarized in Figure~\ref{fig:flow diagram}.
The sets of nodes operating only in one layer are $\mathcal{N}_{A\setminus B,t}:=\mathcal{N}_{A,t}\setminus \mathcal{N}_{B,t}$ and $\mathcal{N}_{B\setminus A,t}:=\mathcal{N}_{B,t}\setminus \mathcal{N}_{A,t}$.
Nodes that lose exactly one functionality transition from $\mathcal{N}_{AB,t}$ to the appropriate single-layer set, while nodes that lose both functionalities exit the system.
We note that, $\mathcal{N}_{A,t}$, $\mathcal{N}_{B,t}$, and $\mathcal{N}_{AB,t}$ are monotone non-increasing in $t$ and the process terminates when no additional overload-induced removals occur and the sets $\mathcal{N}_{AB,t}$, $\mathcal{N}_{A,t}$, and $\mathcal{N}_{B,t}$ stabilize.
In obtaining the recursive equations, we track the fractional sizes of these sets:
\begin{equation}
n_{A,t}=\frac{|\mathcal{N}_{A,t}|}{N}, \qquad
n_{B,t}=\frac{|\mathcal{N}_{B,t}|}{N}, \qquad
n_{AB,t}=\frac{|\mathcal{N}_{AB,t}|}{N}.
\label{eq:fractional_system_size}
\end{equation}
Following the set definitions we can also write $n_{A,t}=n_{A\setminus B,t}+n_{AB,t}$ and $n_{B,t}=n_{B\setminus A,t}+n_{AB,t}$.

For bookkeeping while deriving the recursive equations, let $\mathcal{F}_{i,t}$ denote the nodes that transition at iteration $t$ from both-layer survival to only layer $i\in\{A,B\}$ (i.e., $\mathcal{N}_{AB,t}\to \mathcal{N}_{i\setminus j,t+1}$ with $j\neq i$).
For instance, $\mathcal{F}_{A,2}$ (resp.\ $\mathcal{F}_{B,2}$) denotes the set of nodes that were functioning in both layers at $t=2$, but at iteration $t=3$, they fail in the other layer but remain functional in layer-$A$ (resp.\ layer-$B$).
Analogous to \eqref{eq:fractional_system_size}, we define the fractional size for $\mathcal{F}_{i,t}$ as:
\begin{equation}
f_{A,t}=\frac{|\mathcal{F}_{A,t}|}{N}, \qquad
f_{B,t}=\frac{|\mathcal{F}_{B,t}|}{N}.
\label{eq:fractional_system_size_f}
\end{equation}

We derive recursive equations describing how the surviving fractions $(n_{AB,t},\,n_{A,t},\,n_{B,t})$ and excess loads per surviving node $(Q_{A,t},\,Q_{B,t})$ evolve following an initial attack, with $(L_A,S_A,L_B,S_B)$ drawn i.i.d.\ across nodes from $p_{L_A S_A L_B S_B}$ and cross-layer influence factors $\beta_A$, $\beta_B$.
Table~\ref{tab:recursive_notation} summarizes the notation used below%
\footnote{Probabilistic statements are defined with respect to the probability measure~$\mathbb{P}$, and the associated expectation operator is denoted by~$\mathbb{E}$. The indicator function of an event~$E$ is written as~$\mathbbm{1}[E]$.} .

\textbf{Initial conditions ($t=0$).}
An initial random attack removes a $p$-fraction of nodes from both layers simultaneously, giving
\begin{equation}
n_{AB,0} = n_{A,0} = n_{B,0} = 1-p, \qquad Q_{A,-1} = Q_{B,-1} = 0.
\end{equation}
Since the attacked nodes are chosen uniformly at random, their type-$A$ load is redistributed equally among the $(1-p)\cdot N$ remaining nodes, so the initial excess loads per surviving nodes are
\begin{equation}
Q_{A,0} = \frac{p\,\mathbb{E}[L_A]}{1-p}, \qquad Q_{B,0} = \frac{p\,\mathbb{E}[L_B]}{1-p}.
\label{eq:initial_Q}
\end{equation}
A key feature of the partial-functionality model is that a node can lose functionality in one layer while remaining operational in the other, so the two layers can progress through the cascade at different rates.
The analysis must therefore track transitions among three functional states separately: both-layer survival, $A$-only survival, $B$-only survival, with corresponding set definitions $\mathcal{N}_{AB}$,$\mathcal{N}_{A\setminus B}$ and, $\mathcal{N}_{B\setminus A}$, respectively in Figure~\ref{fig:flow diagram}.

\textbf{Evolution of $n_{AB,t}$.}
At $t=1$, the nodes that survive in both layers are those that satisfy the survival conditions~\eqref{eq:survival_short_A}--\eqref{eq:survival_short_B} under the initial excess loads $(Q'_{A,0},Q'_{B,0})$:
\begin{equation}
n_{AB,1} = n_{AB,0}\cdot\mathbb{P}\!\left[S_A>Q'_{A,0},\,S_B>Q'_{B,0}\right].
\label{eq:n_AB_t1}
\end{equation}
At $t=2$, the surviving pool $\mathcal{N}_{AB,2}\subseteq\mathcal{N}_{AB,1}$ consists of nodes that satisfy the updated excess loads $(Q'_{A,1},Q'_{B,1})$.
Since any node in $\mathcal{N}_{AB,1}$ has already survived the excess loads $(Q'_{A,0},Q'_{B,0})$, the survival probability is conditioned on this event:
\begin{equation}
n_{AB,2} = n_{AB,1} \cdot \mathbb{P}\!\left(\begin{array}{c|c} S_A > Q'_{A,1} & S_A > Q'_{A,0} \\ S_B > Q'_{B,1} & S_B > Q'_{B,0} \end{array}\right).
\label{eq:n_AB_t2}
\end{equation}
The same reasoning applies at every step. In general, for $t=1,2,\ldots$,
\begin{equation} 
n_{AB,t+1} = n_{AB,t} \cdot \mathbb{P}\!\left(\begin{array}{c|c} S_A > Q'_{A,t} & S_A > Q'_{A,t-1} \\ S_B > Q'_{B,t} & S_B > Q'_{B,t-1} \end{array}\right),
\label{eq_main:final_size_v1}  
\end{equation}
with the convention $Q'_{A,-1}=Q'_{B,-1}=0$.
Since $Q'_{A,t}$ and $Q'_{B,t}$ are nondecreasing in $t$ by definition, each conditional probability in~\eqref{eq_main:final_size_v1} equals the ratio $\mathbb{P}[S_A>Q'_{A,t},S_B>Q'_{B,t}]/\mathbb{P}[S_A>Q'_{A,t-1},S_B>Q'_{B,t-1}]$.
Writing out the recursion for $t=1,2,\ldots$ and multiplying the resulting identities gives the telescoping product
\begin{align*}
n_{AB,1} &= n_{AB,0}\cdot\mathbb{P}\!\left[S_A>Q'_{A,0},\,S_B>Q'_{B,0}\right],\\
n_{AB,2} &= n_{AB,1}\cdot\frac{\mathbb{P}\!\left[S_A>Q'_{A,1},\,S_B>Q'_{B,1}\right]}{\mathbb{P}\!\left[S_A>Q'_{A,0},\,S_B>Q'_{B,0}\right]},\\[-1mm]
&\;\vdots\\[-1mm]
n_{AB,t} &= n_{AB,t-1}\cdot\frac{\mathbb{P}\!\left[S_A>Q'_{A,t-1},\,S_B>Q'_{B,t-1}\right]}{\mathbb{P}\!\left[S_A>Q'_{A,t-2},\,S_B>Q'_{B,t-2}\right]}.
\end{align*}
The intermediate factors cancel, yielding
\begin{equation}
\label{eq:AB_recursion}
n_{AB,t}=n_{AB,0}\cdot\mathbb{P}\!\left[S_A>Q'_{A,t-1},\,S_B>Q'_{B,t-1}\right].
\end{equation}

\begin{table}
\caption{\label{tab:recursive_notation}
Notation used in the recursive analysis.
}
\begin{ruledtabular}
\begin{tabular}{ll}
Symbol & Meaning \\
\hline
$n_{AB,t}$ &
\parbox[t]{0.35\textwidth}{Fraction of nodes functioning in both layers at iteration $t$} \\[0.6em]
\\
$n_{A,t}, n_{B,t}$ &
\parbox[t]{0.35\textwidth}{Fractions of nodes functioning in layers $A$ and $B$ at iteration $t$, respectively} \\[0.6em]
\\
$n_{A\setminus B,t}, n_{B\setminus A,t}$ &
\parbox[t]{0.35\textwidth}{Fractions of nodes functioning \textbf{only} in layer-$A$ and \textbf{only} in layer-$B$ at iteration $t$, respectively} \\[0.6em]
\\
$f_{A,t}, f_{B,t}$ &
\parbox[t]{0.35\textwidth}{Fractions of nodes functioning in both layers at iteration $t$ but functioning \textbf{only} in layer-$A$ and \textbf{only} in layer-$B$ at iteration $t+1$, respectively} \\[0.6em]
\\
$Q_{A,t}, Q_{B,t}$ &
\parbox[t]{0.35\textwidth}{Excess loads of type $A$ and type $B$ per surviving node at the end of iteration $t$} \\[0.6em]
\\
$Q'_{A,t}, Q'_{B,t}$ &
\parbox[t]{0.35\textwidth}{Effective excess loads of type $A$ and type $B$ at the end of iteration $t$} \\[0.6em]
\\
$\beta_i$ &
\parbox[t]{0.35\textwidth}{Cross-layer influence factor representing the unit impact of the load in layer $i$ on the overload condition in the other layer, where $i\in\{A,B\}$}
\end{tabular}
\end{ruledtabular}
\end{table}

\textbf{Evolution of $n_{A\setminus B,t}$ and $n_{B\setminus A,t}$.}
The calculation of fractional sizes of nodes functioning in single-layer involves two sources.
For iteration $t+1$, nodes surviving only in layer-$A$ either (i) already were functioning only in $A$ at $t$ ($v_x \in \mathcal{N}_{A\setminus B,t}$) and they remained functional, or (ii) they were functioning in both layers at $t$ and failed in $B$ at the end of that iteration ($v_x \in \mathcal{F}_{A,t}$).
We first characterize the fractional size of the latter, denoted as $f_{A,t}$, and then write the recursion for the fractional size of the nodes only surviving in layer-$A$, denoted as $n_{A\setminus B,t}$.
For brevity, we present the expressions for layer-$A$; the formulas for layer-$B$ are obtained symmetrically.

The nodes that fail in layer-$B$ but survive in layer-$A$ at the end of iteration $t$ (i.e., $\mathcal{F}_{A,t}$) satisfy the survival condition in $A$ ($S_A>Q'_{A,t}$) while violating in layer-$B$ ($S_B<Q'_{B,t}$).
Hence, the fractional size at iteration $t$ is
\begin{equation*}
f_{A,t}=n_{AB,t}\cdot\mathbb{P}\left(\begin{array}{c|c} S_A > Q'_{A,t} & S_A > Q'_{A,t-1} \\ S_B < Q'_{B,t} & S_B > Q'_{B,t-1} \end{array}\right).
\end{equation*}

Using~\eqref{eq:AB_recursion} and expanding the conditional probability, the expression simplifies to
\begin{equation}
f_{A,t} = n_{AB,0}\cdot\mathbb{P}\!\left[S_A>Q'_{A,t},\; Q'_{B,t-1}<S_B<Q'_{B,t}\right].
\label{eq:f_A}
\end{equation}
Equation~\eqref{eq:f_A} provides a form that is easy to interpret. 
The factor $n_{AB,0}$ is the fraction of nodes that survive the initial attack, equal to $1-p$.
The joint probability then describes the event that such a node remains functional in layer-$A$, while it fails in layer-$B$ exactly at iteration~$t$.
This maps precisely to the definition of $\mathcal{F}_{A,t}$.  

Now that we have obtained an expression for $f_{A,t}$, we continue with obtaining a recursion for $n_{A\setminus B,t}$. 
At $t=1$, the only nodes surviving solely in layer-$A$ are those that survived the initial attack but failed in layer-$B$ under $(Q'_{A,0},Q'_{B,0})$, so $\mathcal{N}_{A\setminus B,1}=\mathcal{F}_{A,0}$ and $n_{A\setminus B,1}=f_{A,0}$.

At $t=2$, $\mathcal{N}_{A\setminus B,2}$ consists of all nodes in $\mathcal{F}_{A,1}$ together with those in $\mathcal{F}_{A,0}$ that still survive at iteration $t=2$ as illustrated in Figure~\ref{fig:flow diagram}.
Nodes in $\mathcal{F}_{A,0}$ have already failed in layer-$B$, so their load in $B$ no longer consumes capacity in layer-$A$.
Therefore, by~\eqref{eq:survival_short_single}, these nodes survive in layer-$A$ if they satisfy 
$S_A>Q_{A,1}-\beta_B L_B$.
Hence, the fraction of nodes surviving only in layer-$A$ at $t=2$ is
\begin{multline*}
n_{A\setminus B,2} = f_{A,1} \\
+\, f_{A,0}\cdot\mathbb{P}\!\left[S_A>Q_{A,1}-\beta_B L_B\;\middle|\;\substack{S_A>Q'_{A,0} \\ S_B<Q'_{B,0}}\right].
\end{multline*}
Note that, in the second term, the survival probability for nodes in $\mathcal{F}_{A,0}$ is conditioned on the event defining $\mathcal{F}_{A,0}$, i.e., these nodes survived in layer-$A$ and failed in layer-$B$ at $t=0$, which corresponds to $S_A>Q'_{A,0}$ and $S_B<Q'_{B,0}$.

At $t=3$, $\mathcal{N}_{A\setminus B,3}$ includes all of $\mathcal{F}_{A,2}$ plus the subsets of $\mathcal{F}_{A,1}$ and $\mathcal{F}_{A,0}$ that still satisfy the updated excess load of type~$A$ ($Q_{A,2}$).
Thus, $n_{A\setminus B,3}$ can be calculated as: 
\begin{align*}
n_{A\setminus B,3} &= f_{A,2} \\
&\quad+f_{A,1}\cdot\mathbb{P}\!\left[S_A>Q_{A,2}-\beta_B L_B\;\middle|\;\substack{S_A>Q'_{A,1} \\ Q'_{B,0}<S_B<Q'_{B,1}}\right]\\
&\quad+f_{A,0}\cdot\mathbb{P}\!\left[S_A>Q_{A,2}-\beta_B L_B\;\middle|\;\substack{S_A>Q'_{A,0} \\ S_B<Q'_{B,0}}\right].
\end{align*}
This expression reveals the general structure of the recursion. 
The fraction $n_{A\setminus B,t}$ is obtained by summing the contributions from the sets $\mathcal{F}_{A,j}$, $j=0,1,\dots,t-1$. 
Each contribution is weighted by the probability that a node in the corresponding set survives in layer-$A$ at iteration $t$.
The survival condition is the same in each probability term, namely $S_A>Q_{A,t-1}-\beta_B L_B$ which is the single-layer survival condition in \eqref{eq:survival_short_single} accounting for the failure in layer-$B$.
What changes from one term to another is the conditioning event, since each set $\mathcal{F}_{A,j}$ is defined by the iteration in which they transition from $\mathcal{N}_{AB}$ to $\mathcal{N}_{A\setminus B}$ which is denoted as $j$.
Therefore, for general $t$, we can write
\begin{multline*}
n_{A\setminus B,t} = \sum_{j=0}^{t-1} f_{A,j}\\
\cdot\,\mathbb{P}\!\left[S_A>Q_{A,t-1}-\beta_B L_B\;\middle|\;\substack{S_A>Q'_{A,j} \\ Q'_{B,j-1}<S_B<Q'_{B,j}}\right].
\end{multline*}
Note that for $j=t-1$, the conditional probability will be one since $\beta_BL_B$ is positive and this will produce $f_{A,t-1}$ as in the expanded summation. 
Finally, substituting~\eqref{eq:f_A} for $f_{A,j}$ and combining the two probability terms gives
\begin{equation}
\label{eq:A_recursion}
\begin{split}
n_{A\setminus B,t}
&= n_{AB,0}\sum_{j=0}^{t-1}\mathbb{P}\!\Bigg[\begin{array}{c}
         S_A>Q_{A,t-1}-\beta_B L_B, \\
         S_A>Q'_{A,j}, \\
         Q'_{B,j-1}<S_B<Q'_{B,j}
    \end{array}\Bigg].
\end{split}
\end{equation}
The corresponding expression for $n_{B\setminus A,t}$ follows by symmetry:
\begin{equation}
\label{eq:B_recursion}
\begin{split}
n_{B\setminus A,t}
&= n_{AB,0}\sum_{j=0}^{t-1}\mathbb{P}\!\Bigg[\begin{array}{c}
         S_B>Q_{B,t-1}-\beta_A L_A, \\
         S_B>Q'_{B,j}, \\
         Q'_{A,j-1}<S_A<Q'_{A,j}
    \end{array}\Bigg].
\end{split}
\end{equation}
Using $A$-only and $B$-only surviving fractions, the total layer fractions can be calculated as
\begin{align}
\label{eq:A_tot_recursion}
    n_{A,t} &= n_{AB,t} + n_{A\setminus B,t}, \\
\label{eq:B_tot_recursion}
    n_{B,t} &= n_{AB,t} + n_{B\setminus A,t}.
\end{align}

\textbf{Evolution of $Q_{A,t}$ and $Q_{B,t}$.}
We derive the expression for $Q_{A,t}$; the expression for $Q_{B,t}$ follows by symmetry.
Under global redistribution, load released by failed nodes is redistributed equally to the surviving nodes, so the cumulative type-$A$ load (normalized by~$N$) can be decomposed into initial-attack contribution and cascading contribution:
\begin{multline}
n_{A,t}\,Q_{A,t} = p\,\mathbb{E}[L_A] \\
+ \frac{1}{N}(\text{type-}A\text{ load released} \\
\qquad\qquad\text{by cascading failures up to iteration }t).
\label{eq:load_balance}
\end{multline}
The first term on the right-hand side corresponds to the $p$-fraction of nodes removed at $t=0$, each of which releases its initial load.

The cascading contribution is composed of two parts:
(i)~nodes that were functional in both layers and then fail in layer-$A$; and (ii) nodes that have already lost functionality in layer-$B$ at an earlier iteration and later fail in layer-$A$. 
We keep these two cases separate because they are subjected to different survival conditions. Case~(i) is determined by the two-layer survival condition in~\eqref{eq:survival_short_A}, while case~(ii) is determined by the single-layer survival condition that applies after the additional free space in~\eqref{eq:survival_short_single} becomes available.

In case~(i), a node is functional in both layers and fails in layer-$A$ at iteration~$j$. 
Such a node must satisfy the two-layer survival condition at iteration~$j-1$, namely
$Q'_{A,j-1}<S_A$ and $Q'_{B,j-1}<S_B$, but must violate the layer-$A$ condition at iteration~$j$, so that $Q'_{A,j}>S_A$. 
Therefore, the expected type-$A$ load released through this case up to iteration~$t$ is obtained by summing over all possible failure iterations $j<t$:
\begin{equation}
\sum_{j=0}^{t-1} n_{AB,0}\cdot\mathbb{E}\!\left[L_A\,\mathbbm{1}\!\left(\begin{array}{c} Q'_{A,j-1}<S_A<Q'_{A,j}\\ S_B>Q'_{B,j-1}\end{array}\right)\right].
\label{eq:Q_channel1}
\end{equation}

In case~(ii), a node first moves from $\mathcal{N}_{AB}$ to $\mathcal{N}_{A}$ and then fails in layer-$A$ at a later stage of the cascade. 
Let $j$ denote the iteration at which transition from $\mathcal{N}_{AB}$ to $\mathcal{N}_{A}$ occurs. 
With the notation introduced earlier, these nodes belong to $\mathcal{F}_{A,j}$. 
Thus, to compute the type-$A$ load released by iteration~$t$, we sum over the failures from sets $\mathcal{F}_{A,j}$ with $j<t$. 
A node in $\mathcal{F}_{A,j}$ must survive in layer-$A$ at iteration~$j$, while failing in layer-$B$ at that same iteration. 
This gives the conditions $Q'_{A,j}<S_A$ and $Q'_{B,j-1}<S_B<Q'_{B,j}$. 
After this transition, the node is governed by the single-layer survival condition in layer-$A$. 
Hence, the failure before $t$ would be implied by $S_A<Q_{A,t-1}-\beta_B L_B$. 
Then the corresponding released type-$A$ load from the single-layer failures can be calculated as:
\begin{equation}
\sum_{j=0}^{t-2} n_{AB,0}\cdot\mathbb{E}\!\left[L_A\,\mathbbm{1}\!\left(\begin{array}{c} Q'_{A,j}<S_A<Q_{A,t-1}-\beta_B L_B\\ Q'_{B,j-1}<S_B<Q'_{B,j}\end{array}\right)\right].
\label{eq:Q_channel2}
\end{equation}
The summation stops at $t-2$ rather than $t-1$ because nodes that transition from $\mathcal{N}_{AB}$ to $\mathcal{N}_{A}$ at iteration~$t-1$ cannot fail in layer-$A$ before iteration~$t$.

Combining the initial-attack term with~\eqref{eq:Q_channel1} and~\eqref{eq:Q_channel2} in~\eqref{eq:load_balance}, then dividing by $n_{A,t}$, gives
\begin{widetext}
\begin{align}
\label{eq:QA_recursion}
    Q_{A,t}&=
    \frac{p\mathbb{E}[L_A]+n_{AB,0} \left(
    \sum_{j=0}^{t-1}
    \mathbb{E}\!\left[L_A  \mathbbm{1}\!\left[\begin{array}{c}
        Q'_{A,j-1}<S_A<Q'_{A,j}\\  
        Q'_{B,j-1}<S_B
    \end{array}\right] \right] + \sum_{j=0}^{t-2} \mathbb{E}\!\left[L_A \mathbbm{1}\!\left[\begin{array}{c}
        Q'_{A,j}<S_A<Q_{A,t-1}-\beta_BL_B   \\
        Q'_{B,j-1}<S_B<Q'_{B,j} \end{array}\right]\right]  \right)}{n_{A,t}},\\
\label{eq:QB_recursion}
    Q_{B,t}&=
    \frac{p\mathbb{E}[L_B]+n_{AB,0} \left(
    \sum_{j=0}^{t-1}
    \mathbb{E}\!\left[L_B  \mathbbm{1}\!\left[\begin{array}{c}
        Q'_{B,j-1}<S_B<Q'_{B,j}\\  
        Q'_{A,j-1}<S_A
    \end{array}\right] \right] + \sum_{j=0}^{t-2} \mathbb{E}\!\left[L_B \mathbbm{1}\!\left[\begin{array}{c}
        Q'_{B,j}<S_B<Q_{B,t-1}-\beta_AL_A   \\
        Q'_{A,j-1}<S_A<Q'_{A,j} \end{array}\right]\right]  \right)}{n_{B,t}}.
\end{align}
\end{widetext}
We can further eliminate $n_{A,t}$ and $n_{B,t}$ by substituting~\eqref{eq:A_tot_recursion}-\eqref{eq:B_tot_recursion} together with~\eqref{eq:AB_recursion}-\eqref{eq:B_recursion} into~\eqref{eq:QA_recursion}.
Using $p\,\mathbb{E}[L_A]=n_{AB,0}\,Q_{A,0}$ (and analogously for layer-$B$), the mean-field cascade can be written entirely in terms of $(Q_{A,t},Q_{B,t})$:
\begin{widetext}
\begin{align}
\label{eq:QA_recursion_Qonly}
    Q_{A,t}&=
    \frac{Q_{A,0}+\sum_{j=0}^{t-1}
    \mathbb{E}\!\left[L_A  \mathbbm{1}\!\left[\begin{array}{c}
        Q'_{A,j-1}<S_A<Q'_{A,j}\\  
        Q'_{B,j-1}<S_B
    \end{array}\right] \right] + \sum_{j=0}^{t-2} \mathbb{E}\!\left[L_A \mathbbm{1}\!\left[\begin{array}{c}
        Q'_{A,j}<S_A<Q_{A,t-1}-\beta_BL_B   \\
        Q'_{B,j-1}<S_B<Q'_{B,j} \end{array}\right]\right] }{
    \mathbb{P}\!\left[S_A>Q'_{A,t-1},\,S_B>Q'_{B,t-1}\right] + \sum_{j=0}^{t-1}\mathbb{P}\!\Bigg[\begin{array}{c}
         S_A>Q_{A,t-1}-\beta_B L_B, \\
         S_A>Q'_{A,j}, \\
         Q'_{B,j-1}<S_B<Q'_{B,j}
    \end{array}\Bigg]},\\
\label{eq:QB_recursion_Qonly}
    Q_{B,t}&=
    \frac{Q_{B,0}+\sum_{j=0}^{t-1}
    \mathbb{E}\!\left[L_B  \mathbbm{1}\!\left[\begin{array}{c}
        Q'_{B,j-1}<S_B<Q'_{B,j}\\  
        Q'_{A,j-1}<S_A
    \end{array}\right] \right] + \sum_{j=0}^{t-2} \mathbb{E}\!\left[L_B \mathbbm{1}\!\left[\begin{array}{c}
        Q'_{B,j}<S_B<Q_{B,t-1}-\beta_AL_A   \\
        Q'_{A,j-1}<S_A<Q'_{A,j} \end{array}\right]\right] }{
    \mathbb{P}\!\left[S_A>Q'_{A,t-1},\,S_B>Q'_{B,t-1}\right] + \sum_{j=0}^{t-1}\mathbb{P}\!\Bigg[\begin{array}{c}
         S_B>Q_{B,t-1}-\beta_A L_A, \\
         S_B>Q'_{B,j}, \\
         Q'_{A,j-1}<S_A<Q'_{A,j}
    \end{array}\Bigg]}.
\end{align}
\end{widetext}
Equations~\eqref{eq:QA_recursion_Qonly}--\eqref{eq:QB_recursion_Qonly} show that the mean-field dynamics are fully characterized by the pair $(Q_{A,t},Q_{B,t})$.
The two excess loads are mutually coupled at every step through $Q'_{A,t}=Q_{A,t}+\beta_B Q_{B,t}$ and $Q'_{B,t}=Q_{B,t}+\beta_A Q_{A,t}$, and each update depends on the entire prior history $(Q_{A,j},Q_{B,j})_{j<t}$ rather than on the previous step alone, because every summation is evaluated at the effective loads prevailing when the corresponding nodes failed.
Once a node fails in one layer, its load is set to zero and releases some capacity in the surviving layer, adding a load-dependent history that is absent when both layers fail jointly.

To obtain the final system sizes, we iterate~\eqref{eq:QA_recursion_Qonly}--\eqref{eq:QB_recursion_Qonly} from $t=1$ until $Q_{A,t}$ and $Q_{B,t}$ stabilize, namely until $Q_{A,t}=Q_{A,t+1}$ (or $Q_{A,t}=\infty$) and $Q_{B,t}=Q_{B,t+1}$ (or $Q_{B,t}=\infty$).
If, at any iteration, the denominator in the update for one layer becomes zero, we set the corresponding excess load to infinity.
Thus, $Q_{A,t}=\infty$ ($Q_{B,t}=\infty$) indicates complete collapse of layer-$A$ (layer-$B$).
Let $t^*$ denote the first iteration at which the excess loads stabilize.
Then evaluating~\eqref{eq:AB_recursion}--\eqref{eq:B_tot_recursion} with $(Q_{A,t^*},Q_{B,t^*})$ gives the final fractions $n_{AB,\infty}$, $n_{A,\infty}$, and $n_{B,\infty}$.

Unlike the single-layer flow-network~\cite{single_flow_optimizing} and the joint-functionality multiplex flow  network ~\cite{irsoy2025} cases, the partial-functionality recursion does not, in general, admit a closed-form expression for the excess loads at step~$t$ solely in terms of the initial parameters: the attack size~$p$, the joint load-free-space distribution~$p_{L_A S_A L_B S_B}$, and the cross-layer influence factors~$(\beta_A,\beta_B)$.
This is similar to the interdependent flow-redistribution model of Zhang et al.~\cite{zhang_two_flow_redistribution}, where load transfer across layers makes each failure stage depend on the excess loads prevailing at that stage, so the computation must track which nodes fail at each stage $j$ separately rather than relying only on the current state.
In the partial-functionality setting, a further complication arises: once a node fails in layer-$B$ at step $j$, its survival condition in layer-$A$ shifts to~\eqref{eq:survival_short_single}, $S_A > Q_{A,t-1} - \beta_B L_B$, where the threshold now depends on the node's own load $L_B$ rather than a single shared quantity.
The pair $(Q_A,Q_B)$ therefore cannot be reduced to a fixed-point equation in two scalar excess loads alone.
The final system sizes are found by iterating the recursion until convergence; since each step requires only evaluating population-level expectations, the computation is fast and makes it practical to scan the full $(p, \beta_A, \beta_B)$ parameter space analytically, as we do for the phase diagrams in Section~\ref{sec:cascade_outcome_regimes}.

\section{Numerical Results}
\label{sec:numerical results}

In this section, we present numerical results that validate our analysis and explore how system parameters affect robustness.
Section~\ref{sec:numerical_validation} verifies the recursive equations against Monte Carlo simulations and presents the \textit{robustness curves}, i.e. the variation of final system sizes in each layer as attack size increases, for two representative configurations.
Section~\ref{sec:numerical_beta} studies a counter-intuitive effect of cross-layer influence: increasing the attack size can, in some cases, leave more nodes functional in one layer.
Section~\ref{sec:cascade_outcome_regimes} examines the full cascade-outcome phase diagram and shows how the cross-layer influence $\beta$ moves the boundaries between both-layer survival, single-layer survival, and complete collapse.

To perform numerical simulations, we use different initial load-free space distributions from the Uniform, Pareto, and Weibull families. 
The uniform distribution is denoted by $U(U_{\text{min}}, U_{\text{max}})$, where $U_{\text{min}}$ and $U_{\text{max}}$ are the bounds.  
The Weibull distribution is denoted by $Wei(W_{\text{min}}, \lambda, k)$, with $W_{\text{min}}$ as the minimum, $\lambda$ as the scale, and $k$ as the shape parameter.  
The Pareto distribution is denoted by $Par(P_{\text{min}}, b)$, where $P_{\text{min}}$ is the minimum and $b$ is the shape parameter.

These distributions represent a range of characteristics commonly observed in real-world systems.  
In particular, the uniform distribution serves as a baseline, modeling scenarios where values are evenly distributed over a fixed interval.  
The Pareto distribution captures heavy-tailed behavior, where a small fraction of nodes may have disproportionately large loads or capacities.  
The Weibull distribution provides flexibility to model various statistical distributions, including the Exponential and Rayleigh, and is widely used to describe workloads in engineered systems such as distributed computing platforms~\cite{cloud_comp_weib}.  
Together, these distribution families support a comprehensive and realistic evaluation of system robustness under diverse operating conditions.

Throughout this section, all simulations are conducted under the \textit{global redistribution} rule with $N=10^6$.
In the figures, markers (triangle, cross, circle) report averages over $50$ simulation runs, while the solid lines show the corresponding analytical results calculated by equations~(\ref{eq:QA_recursion_Qonly})–(\ref{eq:QB_recursion_Qonly}) and (\ref{eq:AB_recursion})-(\ref{eq:B_tot_recursion}).

\subsection{Numerical Validation of Recursive Equations}
\label{sec:numerical_validation}

\begin{figure}
\centering
\subfloat[Configuration 1\label{fig:val_config1}]{%
    \includegraphics[width=0.85\linewidth]{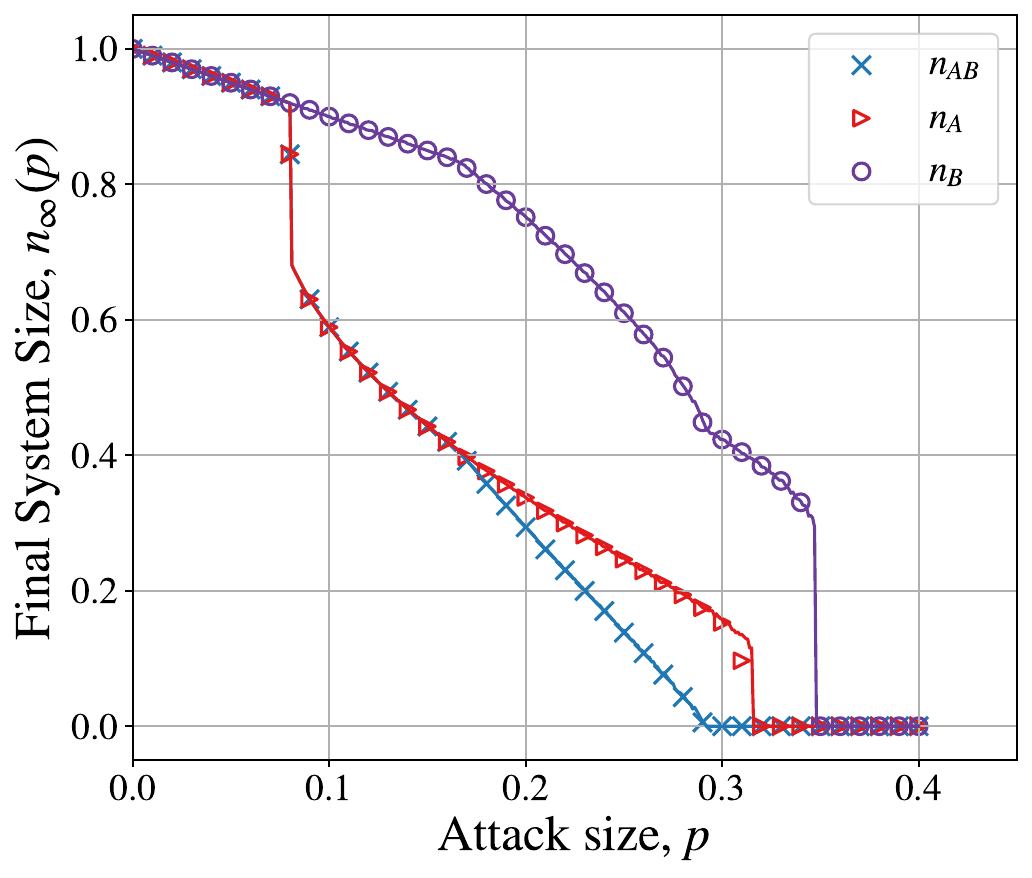}}\\[0.5em]
\subfloat[Configuration 2\label{fig:val_config2}]{%
    \includegraphics[width=0.85\linewidth]{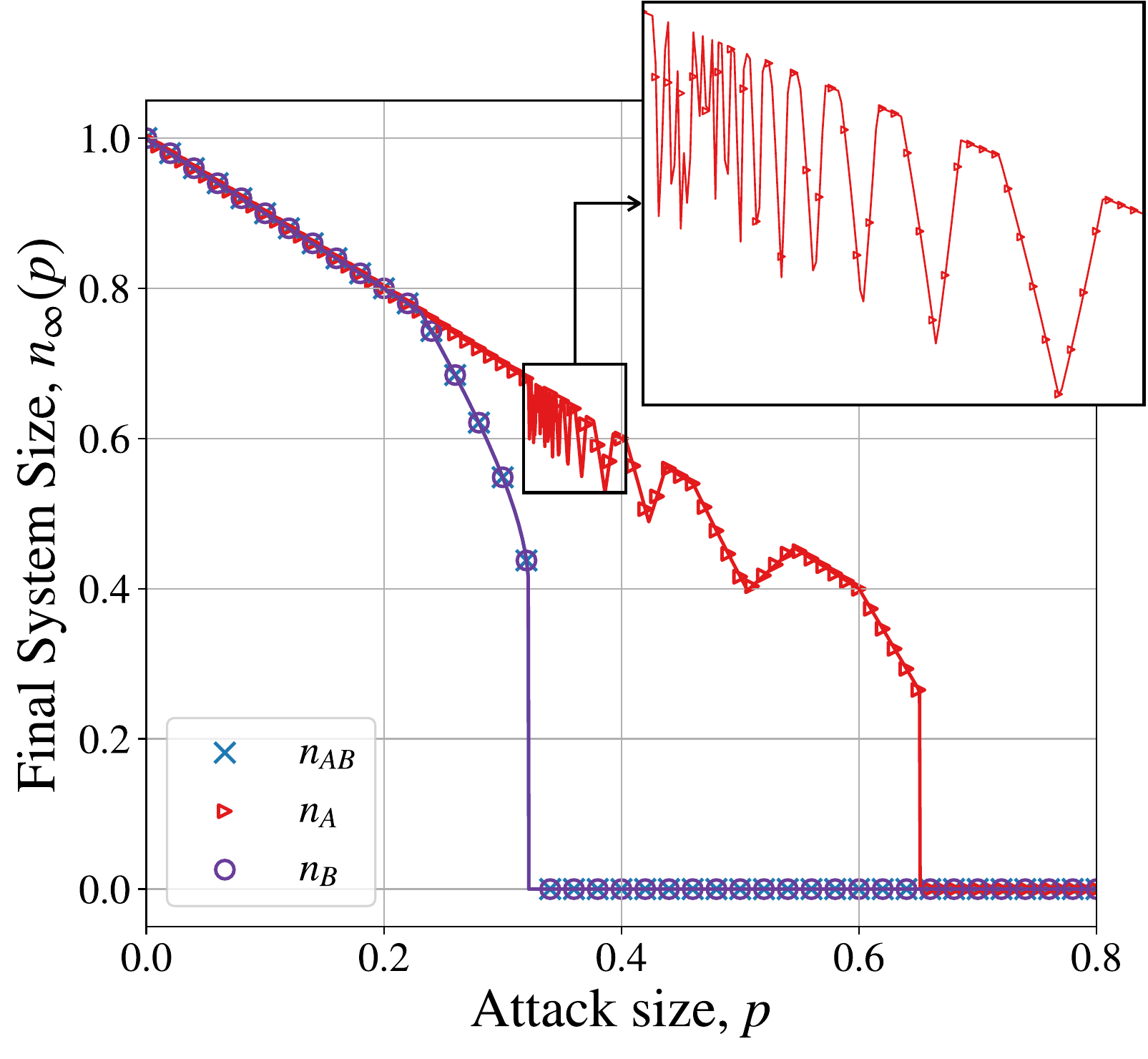}}%
\caption{\label{fig:validation_comparison}
Validation of the recursive equations across two configurations.
In each panel, solid lines show analytical predictions from~(\ref{eq:QA_recursion_Qonly})–(\ref{eq:QB_recursion_Qonly}) and markers report averages over 50 independent Monte Carlo runs with $N=10^6$; red triangles, purple circles, and blue crosses denote layer-$A$, layer-$B$, and both-layer survival fractions, respectively.
(a)~Configuration 1: $L_A\!\sim\!\mathrm{Wei}(10,100,0.4)$, $S_A=3L_A$, $L_B\!\sim\!\mathrm{Wei}(10,10.78,6)$, $S_B\!\sim\!U(20,100)$, $\beta_A=\beta_B=0.1$.
(b)~Configuration 2: $L_A\!\sim\!U(50,100)$, $S_A=3L_A$, $L_B\!\sim\!U(10,40)$, $S_B=3L_B$, $\beta_A=1.0$, $\beta_B=1.0$.
}
\end{figure}

In this subsection, we validate the recursive equations against Monte Carlo simulations.
We tested the recursion under a range of configurations, including loads drawn from Uniform, Pareto, and Weibull families, cases where free space is either sampled independently of the load or assigned proportional to it, and unequal cross-layer influence factors $\beta_A \neq \beta_B$.
Figure~\ref{fig:validation_comparison} shows the fractions of nodes surviving in both layers ($n_{AB}$), in layer-$A$ ($n_A$), and in layer-$B$ ($n_B$) as a function of the initial attack fraction~$p$.
For the analytical results, we evaluate the excess loads from~\eqref{eq:QA_recursion_Qonly}--\eqref{eq:QB_recursion_Qonly} directly until the system reaches a steady state and then evaluate the final system sizes from~\eqref{eq:AB_recursion}--\eqref{eq:B_tot_recursion}. 
For the simulations, we draw node-level loads and free spaces from the same distributions, simulate the cascade until there are no more failures, and report Monte Carlo averages for the final system sizes over $50$ independent runs with $N=10^6$.

Across all tested configurations, the analytical predictions agree closely with simulation averages.
Moreover, we observe some interesting behaviors of the model that are absent in single-layer or joint-functionality cases. 
Figure~\ref{fig:validation_comparison}(a) shows that the system results in distinct final survival states as $p$ increases. 
Around $p \approx 0.09$, nodes functioning in both layers begin to fail in layer-$A$ while remaining operational in layer-$B$; the survival curve for layer-$B$ at this stage closely tracks the linear baseline $1-p$, indicating that these nodes are largely unaffected in layer-$B$ despite losing layer-$A$ functionality. 
Beyond $p \approx 0.18$, a reverse transition sets in where some nodes fail in layer-$B$ but continue operating in layer-$A$. 
In the range $p \in [0.18, 0.3]$, the system settles into a state with coexisting subsets of nodes functioning exclusively in $A$, exclusively in $B$, or in both layers. 
The individual layer transitions to collapse are discontinuous (first-order), suggesting critical threshold behavior, while the decline in $n_{AB}$ is continuous (second-order).
The structure of these distinct survival regimes and how they change with respect to the cross-layer influence factors are examined further in Section~\ref{sec:cascade_outcome_regimes}.

\begin{figure*}[t!]
\centering
\subfloat[$\beta_B=0$\label{fig:beta_b0}]{%
    \includegraphics[width=0.32\textwidth]{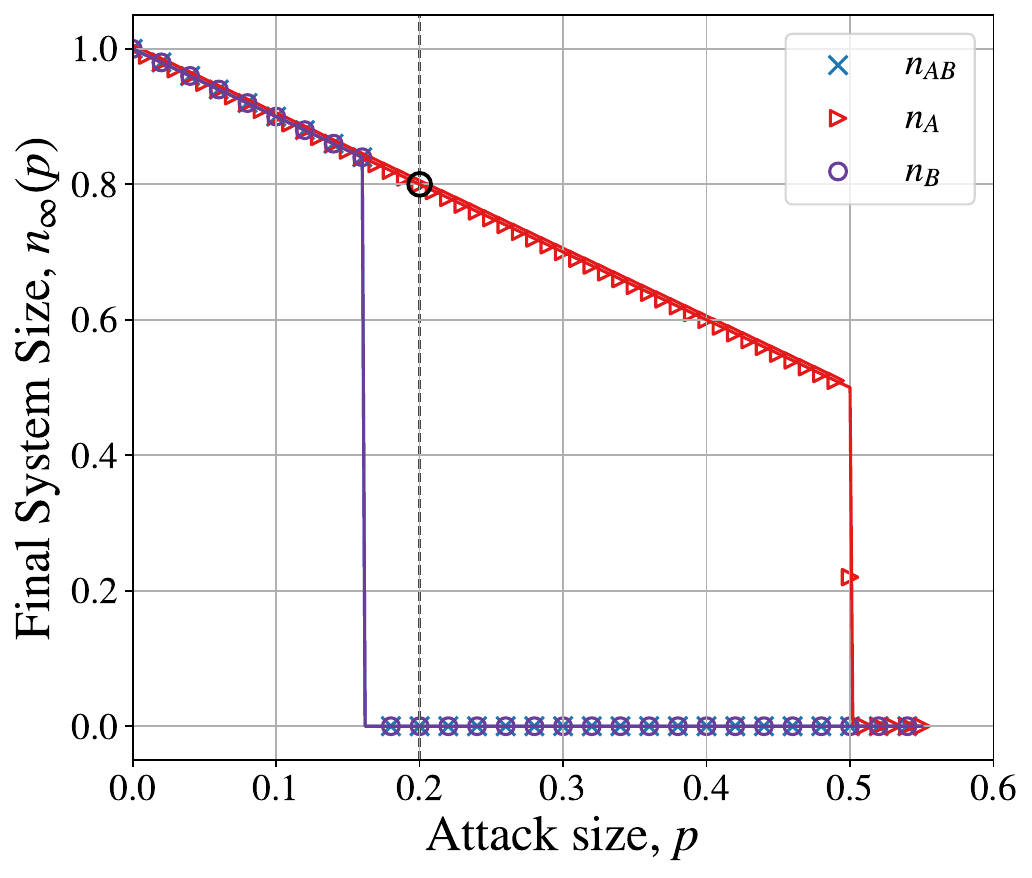}}%
\hfill%
\subfloat[$\beta_B=0.25$\label{fig:beta_b25}]{%
    \includegraphics[width=0.32\textwidth]{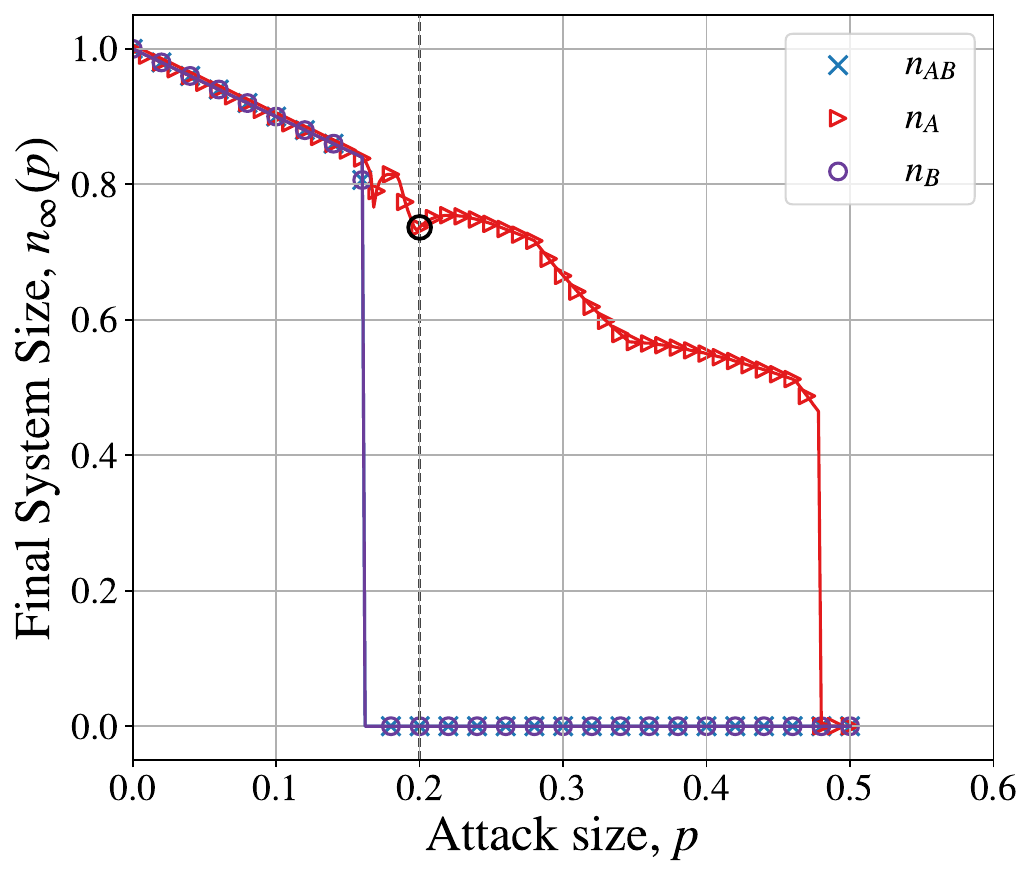}}%
\hfill%
\subfloat[$\beta_B=0.50$\label{fig:beta_b50}]{%
    \includegraphics[width=0.32\textwidth]{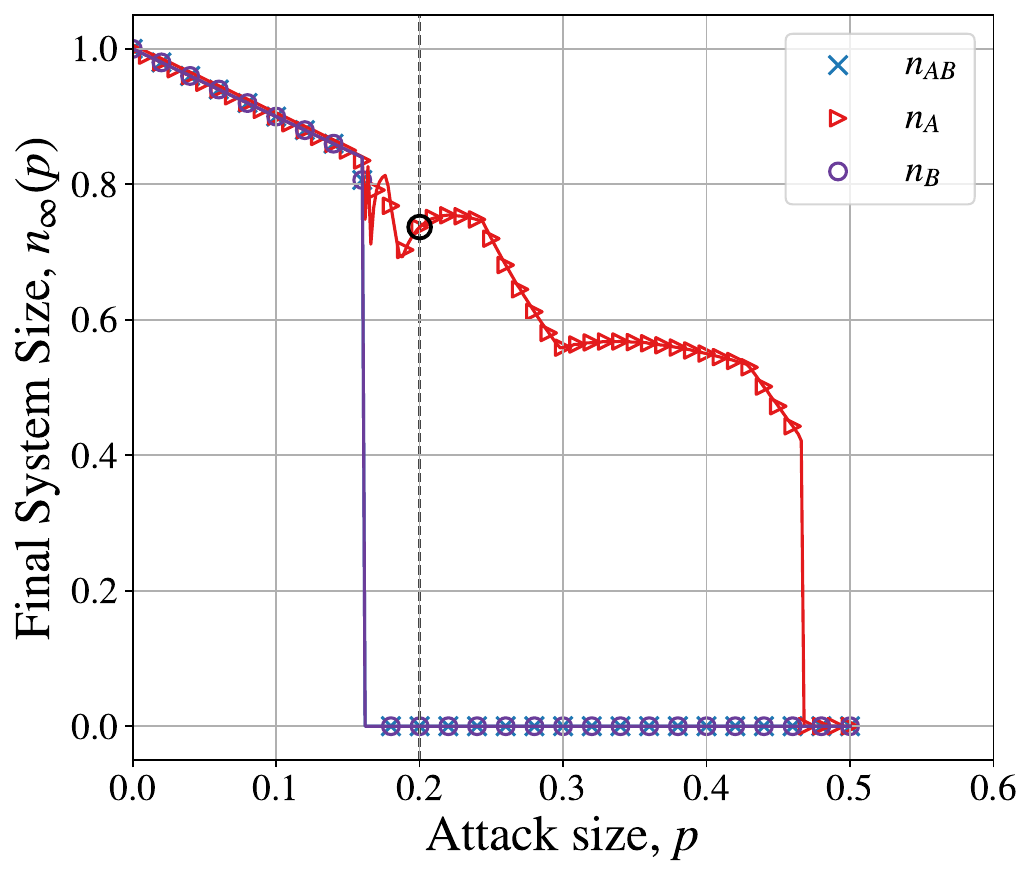}}
\caption{\label{fig:different_beta}
Effect of the cross-layer influence factor $\beta_B$ on cascade outcomes, for fixed $\beta_A=0.20$ and initial distributions $L_A\sim U(10,50)$, $S_A\sim U(30,60)$, $L_B\sim\mathrm{Par}(10,2)$, $S_B=0.5L_B$.
Panels~(a)--(c) show the final system size for nodes surviving in layer-$A$ (red triangles), layer-$B$ (purple circles), and both layers (blue crosses) as a function of the initial attack fraction $p$.
Solid lines are analytical predictions from~(\ref{eq:QA_recursion_Qonly})–(\ref{eq:QB_recursion_Qonly}) and markers are averages over 50 independent Monte Carlo runs with $N=10^6$.
}
\end{figure*}

Figure~\ref{fig:validation_comparison}(b) highlights another interesting consequence of partial functionality.
Once layer-$B$ has collapsed ($n_{B,\infty}=0$), the layer-$A$ curve $n_{A,\infty}(p)$ exhibits a non-monotone behavior with respect to~$p$: it dips near the onset of total $B$ failure and then rises over an intermediate range of~$p$, with small oscillations through the transition.
This is counter-intuitive from a one-layer robustness perspective: a larger attack can leave \emph{more} nodes functioning in layer-$A$.

The basic reason is that failure in layer-$B$ has two competing effects on layer-$A$.
While layer-$B$ is still functioning, its excess load increases the effective load on nodes that are still functioning in both layers, which can trigger additional failures in layer-$A$.
Once layer-$B$ fully collapses, however, its load no longer contributes to the layer-$A$ overload condition for nodes that survive only in layer-$A$.
Thus, a larger attack can sometimes make layer-$B$ collapse earlier and prevent some layer-$A$ nodes from being exposed to the largest layer-$B$ overloads.

This timing effect becomes most visible near the collapse of layer-$B$.
In this configuration, layer-$B$ is more fragile than layer-$A$.
As layer-$B$ approaches collapse, its failed load is redistributed among a rapidly shrinking set of surviving layer-$B$ nodes, which causes the per-survivor excess load $Q_B$ to increase sharply.
Because the layer-$A$ survival condition contains the cross-layer term $\beta_B Q_B$, nodes that are still active in both layers during these late cascade iterations can experience a large additional burden and may fail in layer-$A$ as well.
For some attack sizes, layer-$B$ survives for several additional redistribution rounds, so many both-layer nodes remain exposed to the late-stage spike in $Q_B$ and are lost from layer-$A$.
For a slightly larger attack, layer-$B$ may collapse earlier, before as many both-layer nodes are exposed to the largest values of $Q_B$.
More nodes can then survive as layer-$A$-only nodes, leading to a larger value of $n_{A,\infty}$ despite the larger initial attack.
Thus, small changes in~$p$ can alter the last few cascade iterations and produce the visible oscillations in $n_{AB,\infty}$ and $n_{A,\infty}$.
Despite these abrupt changes in the final system size, the mean-field approximation captures the same behavior and agrees well with the simulation results.
The inset, which uses a finer grid of attack sizes, confirms that the oscillations are also present in the simulations.
We examine how these oscillations change with respect to $\beta_B$ in more detail in the following section.

\subsection{Effect of Cross-Layer Influence}
\label{sec:numerical_beta}

The non-monotone behavior observed in Figure~\ref{fig:validation_comparison}(b) suggests that the timing of failures in one layer can affect the survival of the other layer.
In the partial-functionality model, this effect is controlled by the cross-layer influence factors.
Here, we isolate this mechanism by varying $\beta_B$, which determines how strongly the excess load in layer-$B$ influences the survival condition of layer-$A$, while keeping $\beta_A=0.20$ fixed.
This allows us to test whether the non-monotone layer-$A$ response is caused by the cross-layer influence from layer-$B$ to layer-$A$, and how its strength changes the cascade outcome.

Figure~\ref{fig:different_beta}~(a)-(c) compare three values of $\beta_B$.
When $\beta_B=0$ in Figure~\ref{fig:different_beta}~(a), the excess load in layer-$B$ does not affect the layer-$A$ survival condition, and the non-monotone behavior is absent.
This confirms that the effect is driven by cross-layer influence from layer-$B$ to layer-$A$.
As $\beta_B$ increases, layer-$A$ becomes more exposed to the excess load in layer-$B$, making the transition more pronounced.
This is visible in Figure~\ref{fig:different_beta}~(b)-(c), where the oscillations become larger and the critical attack size decreases as $\beta_B$ increases from $0.25$ to $0.50$.

\begin{figure}[h]
\centering
\includegraphics[width=\linewidth]{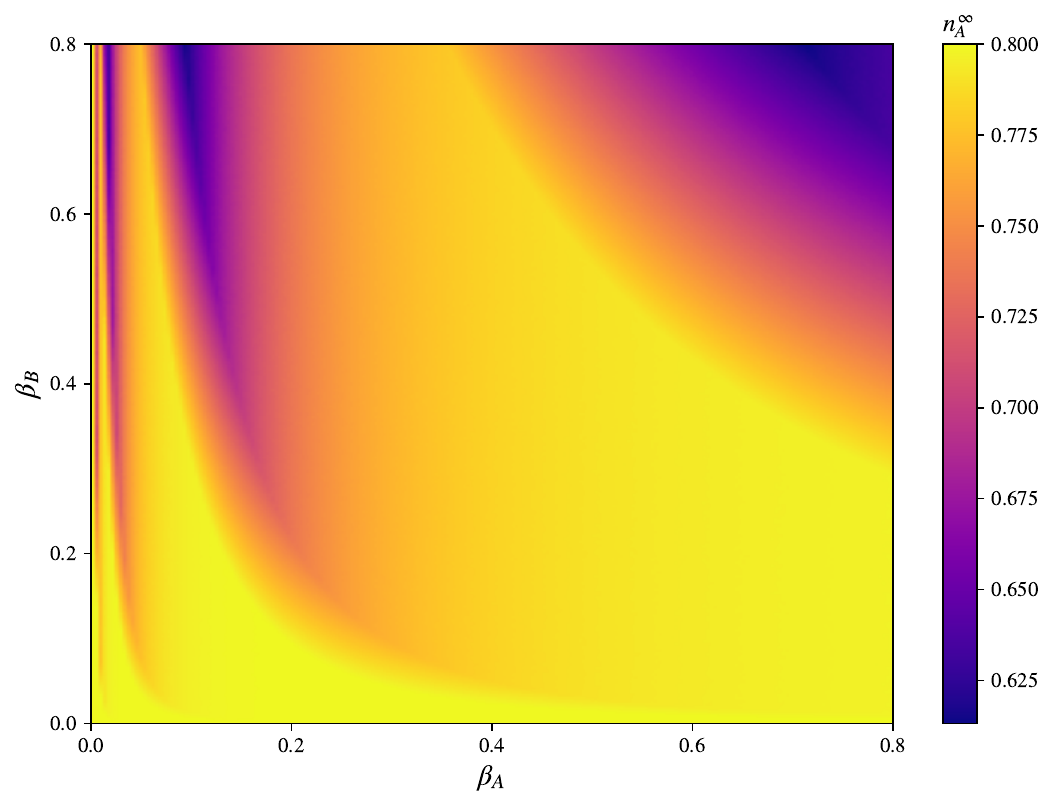}
\caption{\label{fig:beta_heatmap}
Layer-$A$ final size $n_{A,\infty}$ over the $(\beta_A,\beta_B)$ plane at $p=0.20$, computed analytically from~(\ref{eq:QA_recursion_Qonly})–(\ref{eq:QB_recursion_Qonly}) for $L_A\sim U(10,50)$, $S_A\sim U(30,60)$, $L_B\sim\mathrm{Par}(10,2)$, and $S_B=0.5L_B$.
}
\end{figure}

Figure~\ref{fig:beta_heatmap} extends the comparison in Figure~\ref{fig:different_beta} by showing how the final size of layer-$A$ changes when both cross-layer influence factors are varied simultaneously, at the fixed attack size $p=0.20$ marked by the dashed line in Figure~\ref{fig:different_beta}.
The heatmap shows that the effect of cross-layer influence cannot be described only through the strength of one parameter.
Although $\beta_B$ directly influences the layer-$A$ survival condition, $\beta_A$ also affects the outcome by changing how fast the cascade progresses in layer-$B$.
As a result, the final size of layer-$A$ depends on the combined effect of the two cross-layer influence factors.

Figure~\ref{fig:beta_heatmap} maps $n_{A,\infty}$ over the considered $(\beta_A,\beta_B)$ range at $p=0.20$, the attack size marked by the dashed line in Figure~\ref{fig:different_beta}, using the same load and free-space distributions.
Since the configuration has one robust layer, layer-$A$, and one fragile layer, layer-$B$, the two cross-layer influence factors affect $n_{A,\infty}$ in different ways.
The parameter $\beta_B$ enters the layer-$A$ survival condition, so it controls how strongly excess load in layer-$B$ increases the effective load on layer-$A$ and can cause additional failures before layer-$B$ collapses.
The parameter $\beta_A$ affects layer-$A$ more indirectly, through its effect on the survival of layer-$B$.
Values of $n_{A,\infty}$ close to $1-p=0.80$ indicate that layer-$A$ loses almost no additional nodes beyond the initial attack, while lower values correspond to additional layer-$A$ failures caused by the cascade.

Figure~\ref{fig:beta_heatmap} shows that the dependence of $n_{A,\infty}$ on these two parameters, $(\beta_A,\beta_B)$, is not monotone in a simple way.
For small to moderate values of $\beta_B=0-0.1$, layer-$A$ remains highly robust over a wide range of $\beta_A$, as indicated by the broad bright region in the lower part of the figure.
For this region, changing $\beta_A$ has a limited effect on $n_{A,\infty}$ because the cross-layer influence from layer-$B$ to layer-$A$ is still weak.
For larger values of $\beta_B$, the final size of layer-$A$ becomes more sensitive to both cross-layer influence factors.
In particular, the upper-right region becomes darker, showing that strong influence in both directions leads to more additional failures in layer-$A$.
At small $\beta_A$, the narrow alternating bands on the upper left indicate that small changes in the cross-layer influence factors can shift the cascade outcome noticeably, which is consistent with the sharp transitions observed in Figure~\ref{fig:different_beta}.
When the cross-layer influence from the failed layer is high, the final system size in the surviving layer becomes more sensitive to additional losses from cascade timing.
Overall, Figure~\ref{fig:beta_heatmap} shows that layer-$A$ is most vulnerable when the direct burden from layer-$B$ is strong and layer-$B$ participates long enough in the cascade, while it remains close to $1-p$ in the lower-$\beta_B$ region.

\subsection{Cascade Outcome Regimes}
\label{sec:cascade_outcome_regimes}

\begin{figure*}
\centering
\includegraphics[width=\textwidth]{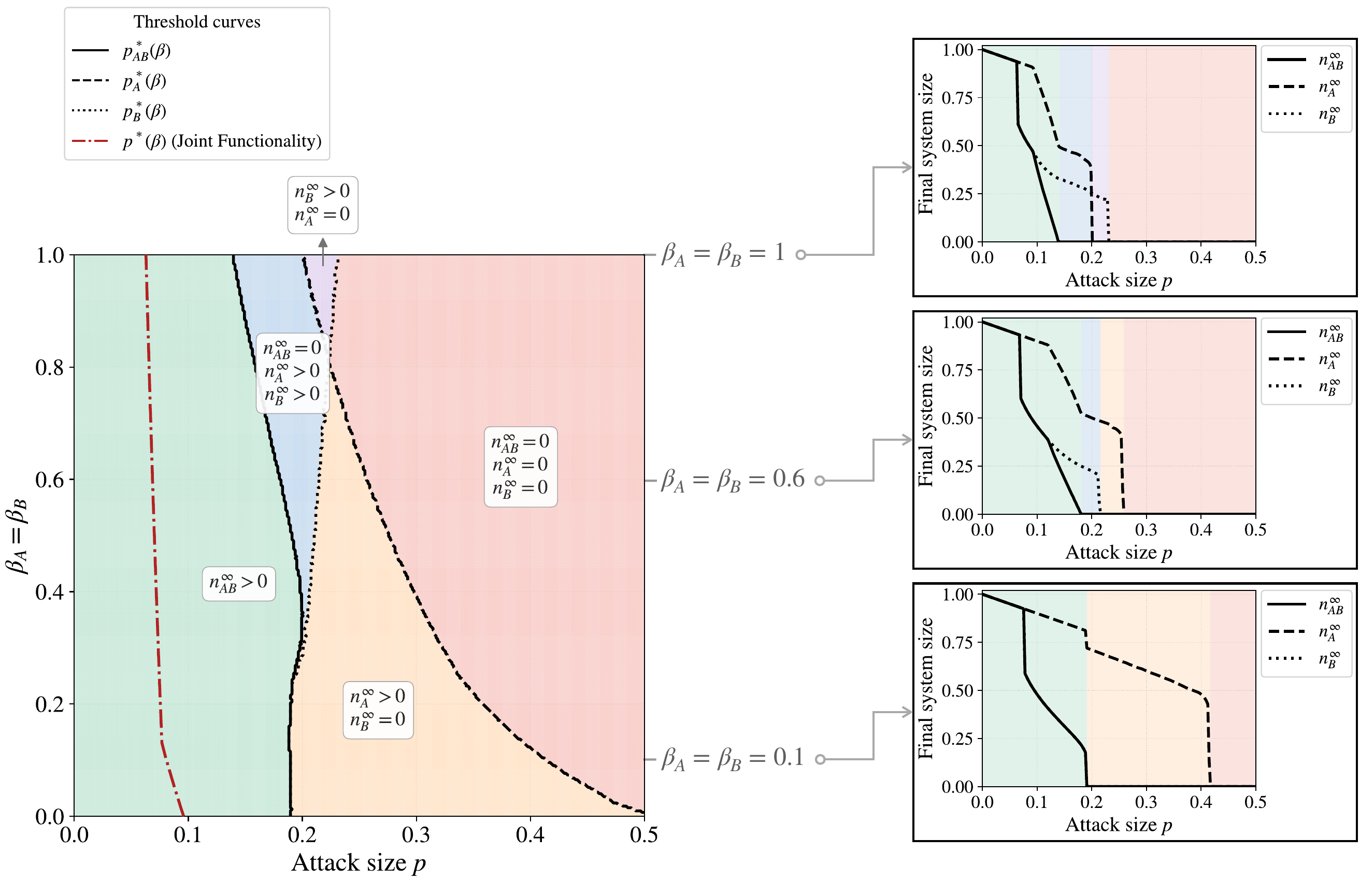}
\caption{\label{fig:phase_diagram}
Cascade-outcome phase diagram in the $(p,\beta)$ plane for the symmetric case $\beta_A=\beta_B=\beta$, computed analytically from~(\ref{eq:QA_recursion_Qonly})–(\ref{eq:QB_recursion_Qonly}) with $L_A\sim U(50,150)$, $S_A=2L_A$, $L_B\sim\mathrm{Wei}(15,100,0.4)$, and $S_B=2L_B$.
The five shaded regions correspond to distinct steady-state cascade outcomes: dual-layer survival ($n_{AB,\infty}>0$, green); individual single-layer survival in both layers but not jointly ($n_{AB,\infty}=0$, $n_{A,\infty}>0$, $n_{B,\infty}>0$, blue); A-only survival ($n_{A,\infty}>0$, $n_{B,\infty}=0$, orange); B-only survival ($n_{B,\infty}>0$, $n_{A,\infty}=0$, teal); and complete collapse (pink).
The solid, dashed, and dotted curves are the critical attack fractions $p^*_{AB}(\beta)$, $p^*_A(\beta)$, and $p^*_B(\beta)$, respectively, as defined in~\eqref{eq:critical_p_a}--\eqref{eq:critical_p_ab}.
The dash--dot curve is $p^*(\beta)$ for the joint-functionality model under the same configuration~\cite{irsoy2025}.
The three right-hand panels show the final system sizes $n_{AB,\infty}$, $n_{A,\infty}$, and $n_{B,\infty}$ as a function of $p$ at $\beta=0.1$, $0.6$, and $1.0$, with background shading matching the phase-diagram regions.
}
\end{figure*}

The previous subsection focused on how cross-layer influence can produce non-monotone changes in the final size of a surviving layer.
We now turn our focus to characterize the possible steady-state outcomes of the cascade more broadly.
Under partial functionality, the system does not have only two possible outcomes, survival or collapse.
Instead, nodes may survive in both layers, only in layer-$A$, only in layer-$B$, or fail in both layers, as described by the sets $N_{AB}$, $N_{A\setminus B}$, and $N_{B\setminus A}$.
This makes it useful to map the parameter ranges where both layers survive, where only single layer remains functional, and where the system completely collapses.

Figure~\ref{fig:phase_diagram} maps the steady-state cascade outcomes across the $(p,\beta)$ plane for the symmetric case $\beta_A=\beta_B=\beta$, computed analytically from recursive equations for the distributions given in the caption. 
The left panel partitions the parameter space into five steady-state regions separated by the critical boundaries $p^*_{AB}(\beta)$, $p^*_A(\beta)$, and $p^*_B(\beta)$ (solid, dashed, and dotted curves, respectively, as defined in~\eqref{eq:critical_p_a}--\eqref{eq:critical_p_ab}). 
The three right-hand panels show cross-sections at $\beta=0.1$, $0.6$, and $1.0$, plotting $n_{AB,\infty}$, $n_{A,\infty}$, and $n_{B,\infty}$ as functions of $p$, with background shading matching the corresponding region in the left panel.

The three cross-sections on the right of Figure~\ref{fig:phase_diagram} reveal how the cascade outcome changes with the cross-layer influence factors. 
For small values, $\beta=0.1$, the loss of dual-layer survival occurs at the same attack size as the collapse of layer-$B$. 
The system then enters the A-only survival region, where $n_{A,\infty}>0$ and $n_{B,\infty}=0$, before reaching complete collapse at a larger attack size. 
For the intermediate value $\beta=0.6$, dual-layer survival is lost before either individual layer collapses. 
This produces a region where both $n_{A,\infty}$ and $n_{B,\infty}$ are positive, but no node remains functional in both layers. 
As $p$ increases further, layer-$B$ collapses first, followed by layer-$A$. 
For stronger cross-layer influence, $\beta=1.0$, the ordering changes: after the loss of dual-layer survival, layer-$A$ collapses before layer-$B$, leading to a B-only survival region before complete collapse.

The resulting phase diagram reveals several important features of the cascade transitions. 
The boundary curves show that increasing $\beta$ does not affect the two layers equally. 
As $\beta$ increases, $p^*_A(\beta)$ shifts toward smaller values of $p$, meaning that layer-$A$ collapses under progressively weaker attacks. 
At the same time, $p^*_B(\beta)$ shifts toward larger values of $p$, meaning that layer-$B$ can withstand larger attacks. 
Thus, stronger cross-layer influence is beneficial for one layer while being detrimental for the other. 
This asymmetry is not immediately apparent from the local failure condition, since the cross-layer influence factors are symmetric. 
It emerges from the interaction between the strength of cross-layer influence and the different load and free-space distributions in the two layers.

The right panels of Figure~\ref{fig:phase_diagram} further illustrate the order and nature of the transitions for different values of $\beta$. 
For small $\beta$, the collapses of $n_{AB,\infty}$ and $n_{B,\infty}$ occur at the same attack size, and both transitions are abrupt (i.e., discontinuous transition). 
As $\beta$ increases, the dual-layer survival threshold separates from the layer-$A$ threshold, and the transition of $n_{AB,\infty}$ to zero happens through a gradual decrease (i.e., continuous transition). 
A similar change is observed for $n_{B,\infty}$: the abrupt drop around $p=0.2$ at smaller $\beta$ turns into a continuous decrease as the cross-layer influence increases. 
These observations show that cross-layer influence affects not only the critical attack sizes, but also the nature of the cascade transitions.

To compare the partial- and joint-functionality models, Figure~\ref{fig:phase_diagram} also includes the dash-dot curve (red) corresponding to the critical attack size for the joint-functionality model under the same load and free-space configuration~\cite{irsoy2025}.
In the joint-functionality model, the system has only two possible outcomes: dual-layer survival or complete collapse.
Therefore, the region to the left of this curve corresponds to dual-layer survival, while the region to the right corresponds to system failure.
In contrast, the partial-functionality model allows intermediate regimes in which one or both layers remain individually functional after dual-layer survival has been lost.
The appearance of the orange, purple, and blue regions is therefore a direct consequence of allowing partial functionality.
More notably, the region of dual-layer survival also expands relative to the joint-functionality threshold.
This suggests that nodes failing in only one layer can still carry and redistribute load in their surviving layer, which reduces the burden on nodes that remain functional in both layers.

\begin{figure}[t]
\centering
\subfloat[Configuration 1\label{fig:phase_config1}]{%
    \includegraphics[height=4cm]{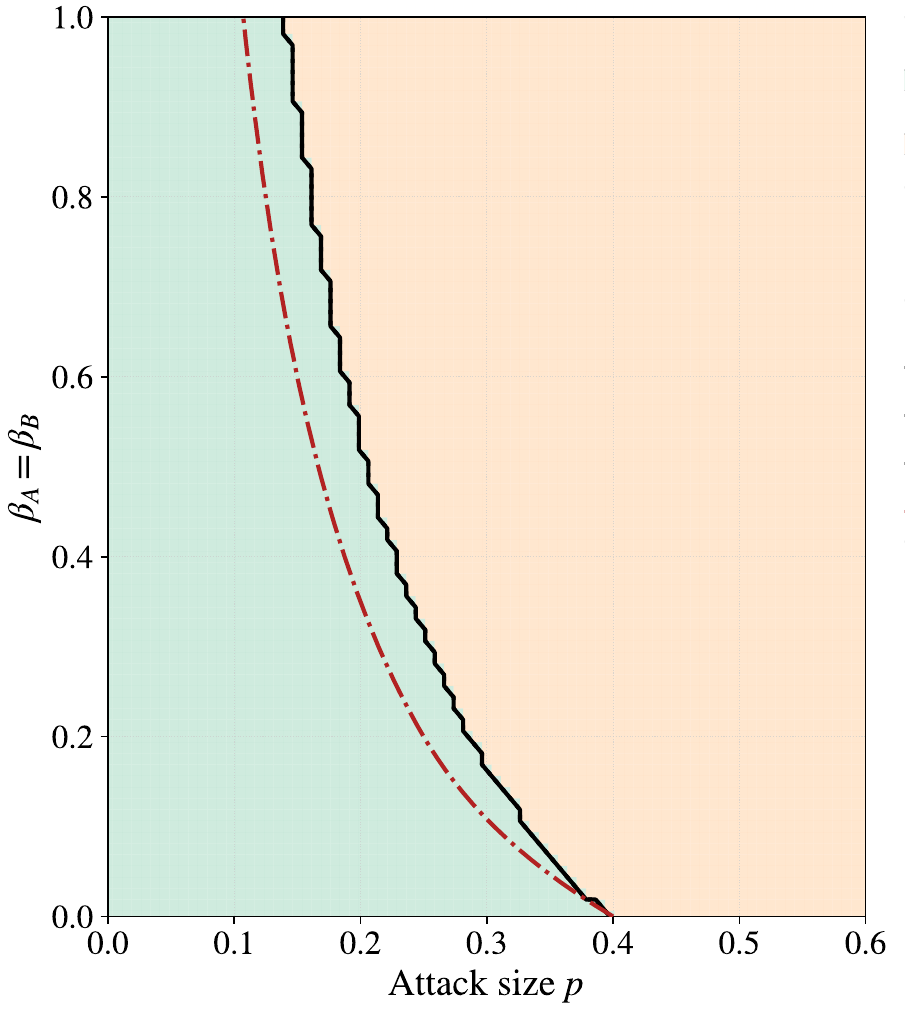}}%
\subfloat[Configuration 2\label{fig:phase_config2}]{%
    \includegraphics[height=4cm]{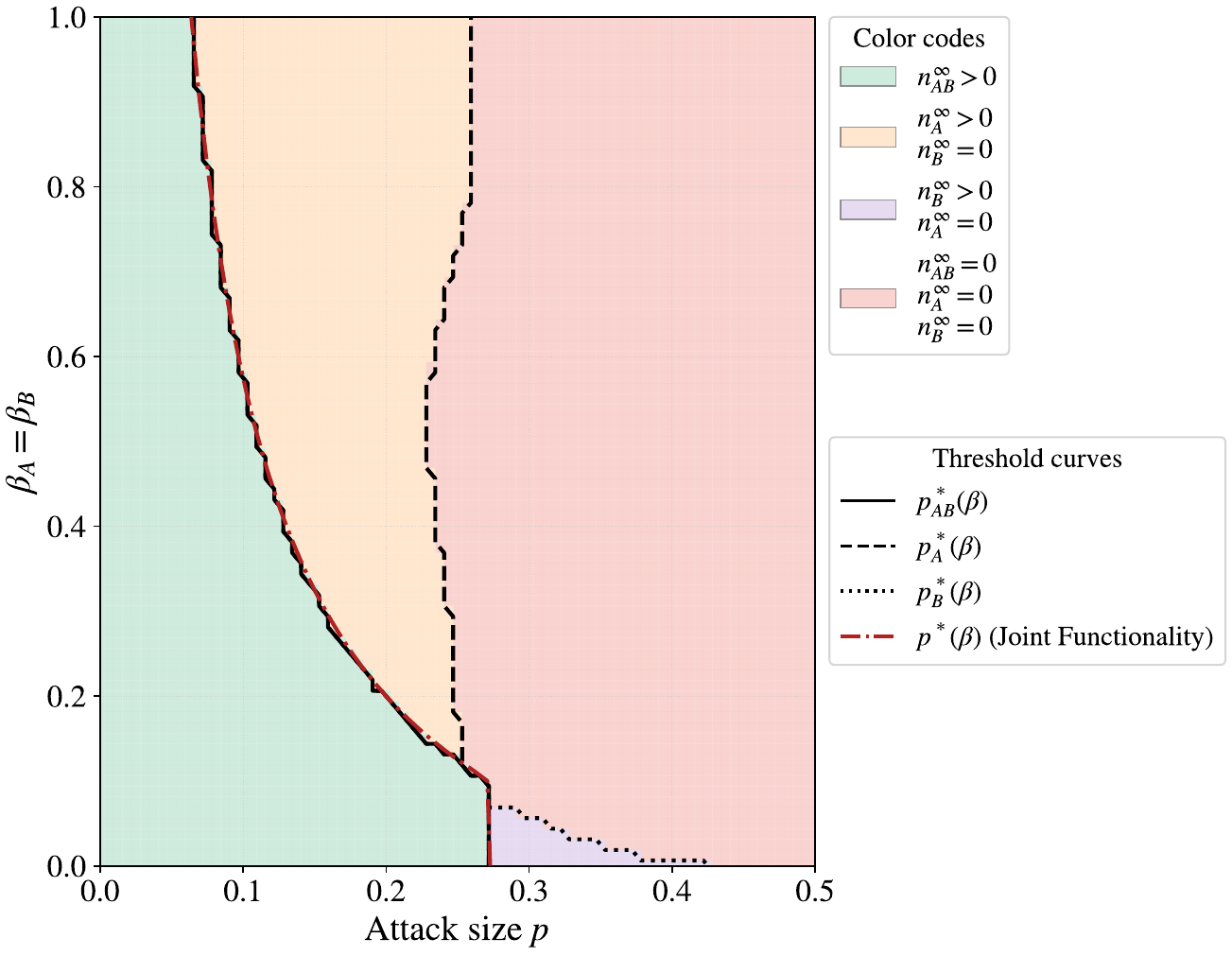}}
\caption{\label{fig:phase_diagram_alt}
Cascade-outcome phase diagrams in the $(p,\beta)$ plane for $\beta_A=\beta_B=\beta$, computed analytically from~(\ref{eq:QA_recursion_Qonly})–(\ref{eq:QB_recursion_Qonly}).
Shading and curve styles follow Figure~\ref{fig:phase_diagram}; the dash--dot curve is $p^*(\beta)$ for the joint-functionality overload condition in Ref.~\cite{irsoy2025}.
(a)~Configuration 1: $L_A\sim U(50,150)$, $S_A=3L_A$, $L_B\sim U(10,40)$, $S_B=1.5L_B$.
(b)~Configuration 2: $L_A\sim\mathrm{Par}(50,2)$, $S_A=0.75L_A$, $L_B\sim\mathrm{Par}(5,2)$, $S_B=1.5L_B$.
}
\end{figure}

Additionally, we repeated the analysis for several other load and free-space configurations; two representative examples are shown in Figure~\ref{fig:phase_diagram_alt}.
Similar steady-state regions arise, but the locations and ordering of $p^*_{AB}(\beta)$, $p^*_A(\beta)$, and $p^*_B(\beta)$ depend strongly on the underlying distributions.
Each panel also includes the joint-functionality threshold $p^*(\beta)$ from Ref.~\cite{irsoy2025} as the dash--dot curve.
In Figure~\ref{fig:phase_diagram_alt}(a), the curves $p^*_{AB}(\beta)$ and $p^*_B(\beta)$ overlap over the parameter range. 
Thus, the loss of dual-layer survival occurs almost simultaneously with the collapse of layer-$B$, and the system mainly transitions from dual-layer survival to A-only survival. 
In this configuration, the joint-functionality threshold lies below the partial-functionality boundary for dual-layer survival, showing a clear robustness gain from allowing nodes to fail in one layer while remaining active in the other. 
In Figure~\ref{fig:phase_diagram_alt}(b), the behavior is different. 
There are distinct ranges of $\beta$ where $p^*_{AB}(\beta)$ overlaps with $p^*_B(\beta)$, where it overlaps with $p^*_A(\beta)$, and where all three thresholds are nearly coincident. 
In contrast to panel~(a), the partial-functionality model does not produce a comparable gain in the dual-survival threshold relative to the joint-functionality case.

Figure~\ref{fig:phase_diagram_alt}(b) also illustrates the non-monotone effect of cross-layer influence factors on layer-$A$ survival. 
As $\beta$ first increases, the A-only survival region shrinks, indicating that cross-layer influence makes layer-$A$ more vulnerable to load coming from layer-$B$. 
For larger $\beta$, this region expands again. 
This reversal is consistent with the capacity relief discussed earlier: when layer-$B$ collapses earlier in the cascade, the cross-layer load pressure on layer-$A$ is removed, which can allow layer-$A$ to survive over a wider range of attack sizes.

Overall, Figures~\ref{fig:phase_diagram} and~\ref{fig:phase_diagram_alt} show that cascade outcomes in the partial-functionality model are shaped by the combined effect of cross-layer influence and the load-free space distributions in each layer. 
The same increase in $\beta$ can shift thresholds, change which layer is more vulnerable, and alter whether partial functionality improves dual-layer robustness. 
These observations highlight why multiple critical thresholds are needed to describe robustness in this setting, rather than a single collapse point for the whole multiplex system.

\section{Improving Robustness Through Free-Space Allocation}
\label{sec:fsa}

\begin{figure*}[t]
\centering
\subfloat[Layer-weighted equal FSA]{%
    \includegraphics[width=0.32\textwidth]{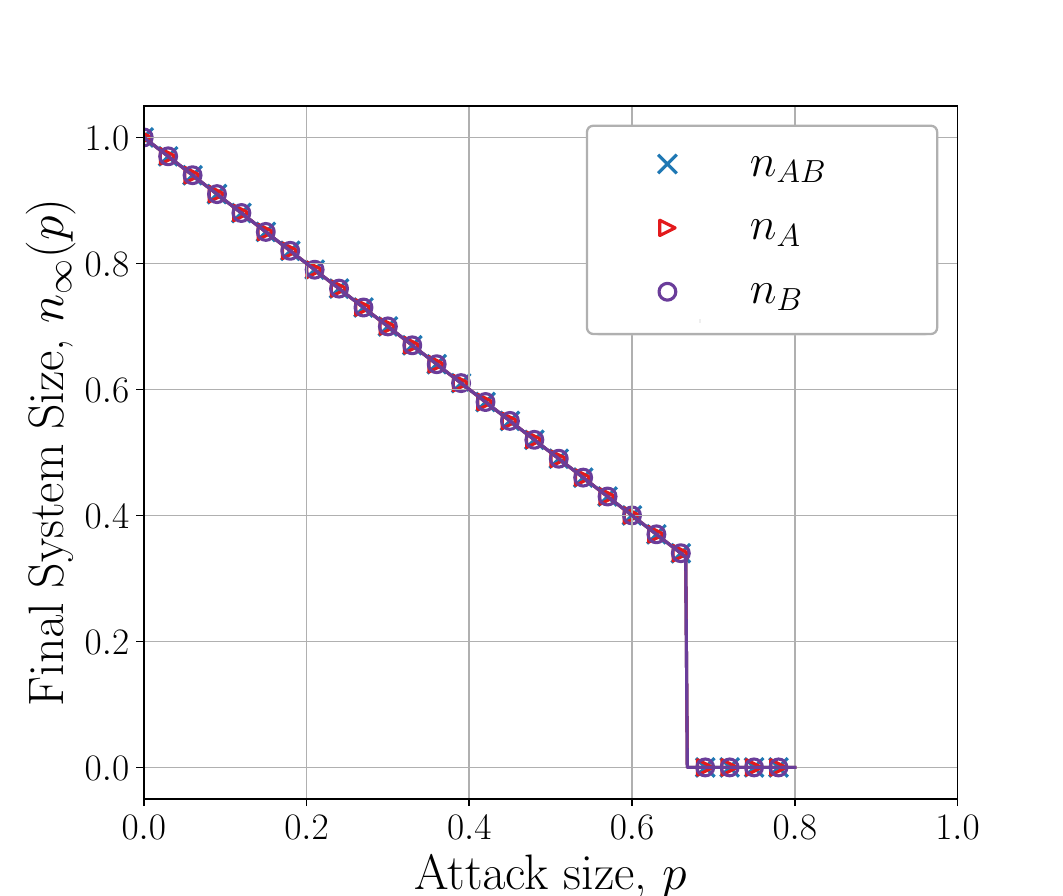}}%
\hfill%
\subfloat[Equal FSA]{%
    \includegraphics[width=0.32\textwidth]{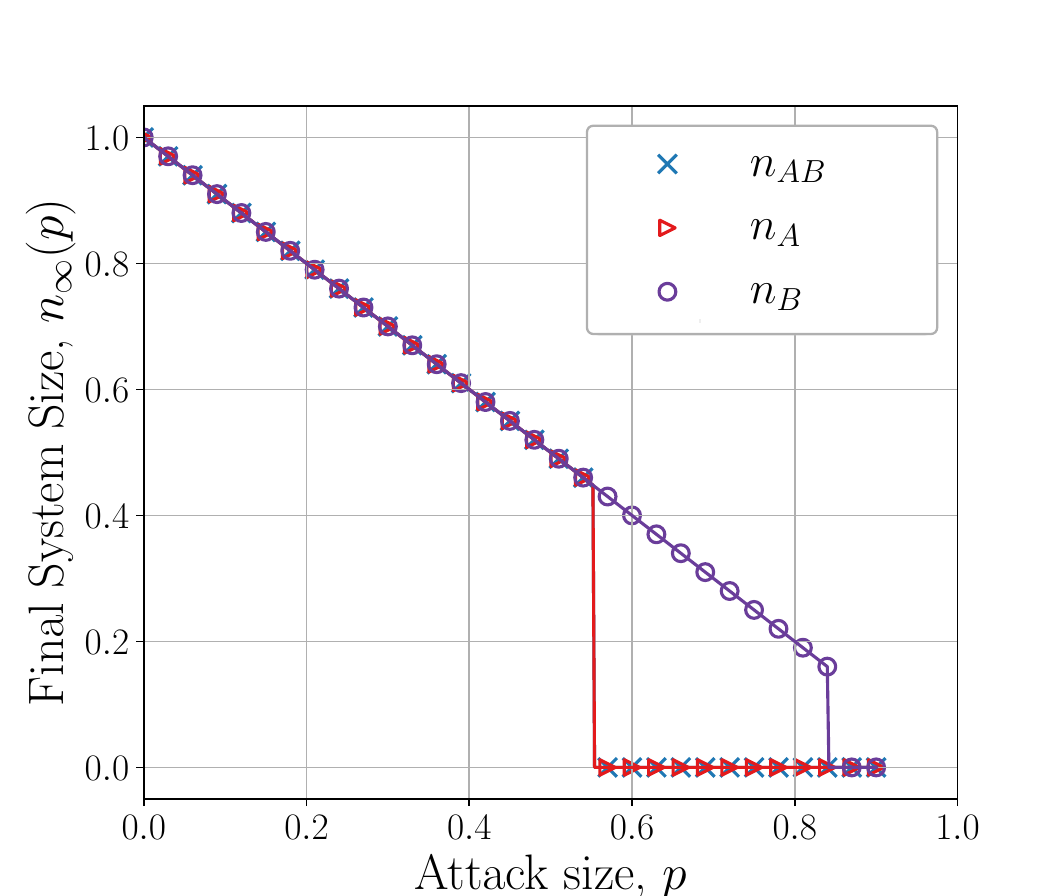}}%
\hfill%
\subfloat[Equal tolerance factor]{%
    \includegraphics[width=0.32\textwidth]{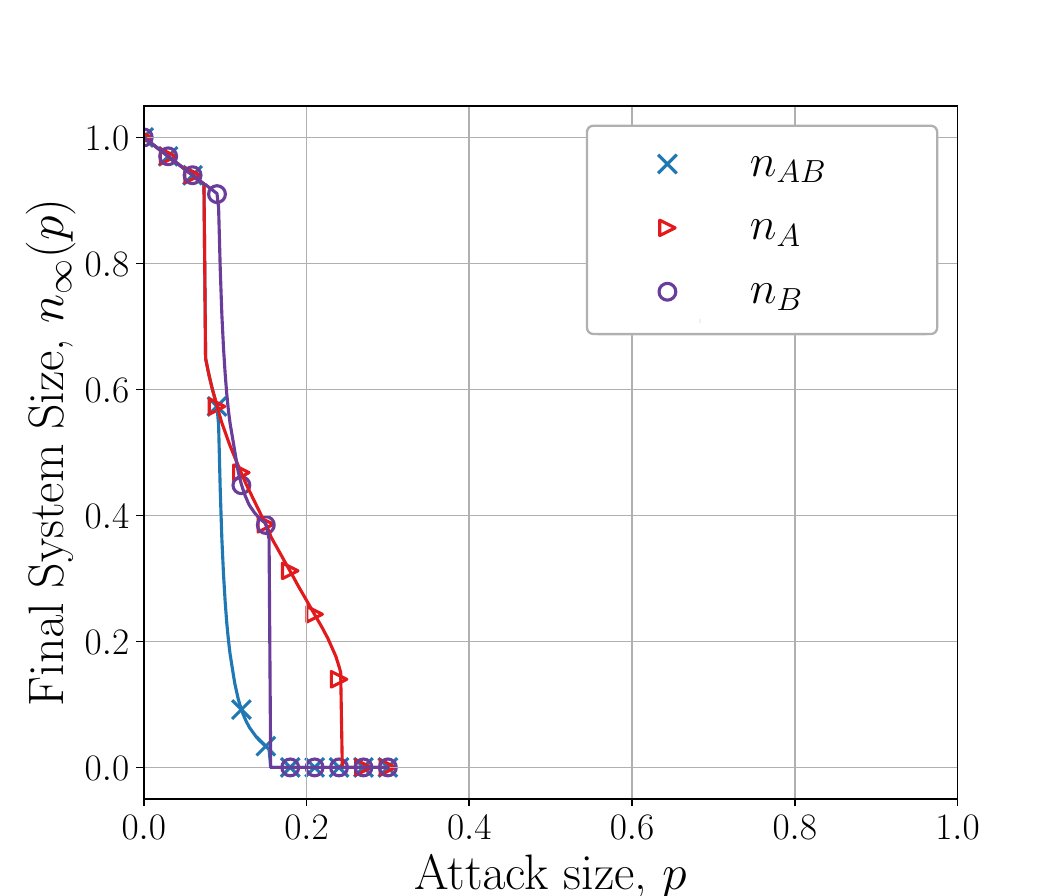}}
\caption{\label{fig:optimal_strategy}
Final system size for nodes surviving in $A$ (red triangle), $B$ (purple circle), and both layers (blue cross), for three free-space allocation strategies:
(a) layer-weighted equal FSA;
(b) equal FSA;
(c) equal tolerance factor.
The initial distributions are $L_A \sim \mathrm{Wei}(10,84.25,0.4)$ and $L_B \sim \mathrm{Par}(5,2)$, with $\beta_A=\beta_B=0.2$.
For equal FSA, each layer receives 360 free space, distributed equally among nodes.
For equal tolerance factor, $\alpha=2.4$ satisfies the total free-space constraint.
For layer-weighted equal FSA, free space is allocated across layers in proportion to expected loads adjusted by the cross-layer influence factors, resulting in $S_A=584$ and $S_B=136$.
}
\end{figure*}

In many motivating applications, the total available free space or capacity is inherently limited.
How this limited capacity is allocated across layers and nodes can play a critical role in mitigating large-scale disruptions.
Section~\ref{sec:numerical_strategy_global} compares three benchmark free-space allocation (FSA) strategies under global redistribution.
Section~\ref{sec:numerical_strategy_local} then evaluates these strategies, together with a proposed local-risk-weighted allocation rule, under the local redistribution rule on explicit network topologies.
For the global-redistribution experiments, we use $N=10^6$ nodes with markers reporting averages over $50$ simulation runs and solid lines showing analytical results obtained from recursive equations as in Section~\ref{sec:cascade_outcome_regimes}.
For the local-redistribution experiments, we use networks of size $N=10^5$ with $50$ replications; since analytical results are not available in this setting, all plotted points represent simulation averages.

\subsection{Comparison under Global Redistribution}
\label{sec:numerical_strategy_global}
We consider settings where the total free space across the two layers is fixed, i.e., 
\(\mathbb{E}[S_A] + \mathbb{E}[S_B] = S_{\text{total}}\), 
where \(S_{\text{total}}\) is constant. 
Under the \textit{partial-functionality overload} condition studied here, characterizing the optimal allocation of this fixed free space is analytically challenging, because the final system size has no closed-form expression and the multiple functional states require separate failure evaluations.   
Given these limitations, we focus on a numerical comparison of three practical \textbf{free-space allocation (FSA)} strategies used in related flow-redistribution models:

\begin{itemize}
    \item[\textbf{i)}] \textbf{Layer-weighted equal FSA:} The total free space is divided between layer-$A$ and layer-$B$ in proportion to their expected loads adjusted by the \textit{cross-layer influence factors}, i.e., in proportion to \( \mathbb{E}[L_A] + \beta_B \mathbb{E}[L_B] \) and \( \mathbb{E}[L_B] + \beta_A \mathbb{E}[L_A] \), respectively. Then, within each layer, the allocated free space is distributed evenly across all nodes. This strategy was shown to be optimal for the multiplex flow network model with \textit{joint functionality}~\cite{irsoy2025}. 
    \item[\textbf{ii)}] \textbf{Equal FSA:} The total free space is split equally between the two layers and distributed evenly across all nodes in each layer. This strategy corresponds to the optimal allocation rule for the single-layer flow network and its variant with fractional load loss~\cite{single_flow_optimizing,ozel_2018}. 
    \item[\textbf{iii)}] \textbf{Equal tolerance factor:} Each node is assigned free space in proportion to its load, i.e., \( S_{x,i} = \alpha L_{x,i} \), where \( \alpha \) is a constant for all \( x \in \mathcal{N} \). This strategy has been used broadly in flow-redistribution models~\cite{scala,lee2016strength,WANG2025130373,pei_et.al.,Wang_Jin_Zhao_2021,Zhou_Elmokashfi_2017}.
\end{itemize}

We evaluated these strategies under various initial load configurations, including Weibull–Uniform, Uniform–Pareto, and Weibull–Pareto distributions.  
As the observed trends were consistent across all configurations, we present results for the Weibull–Pareto case in Figure~\ref{fig:optimal_strategy}, which is representative of the overall behavior.

\begin{figure*}[t]
\centering
\subfloat[ER Network: Wei-Par]{%
    \includegraphics[width=0.31\textwidth]{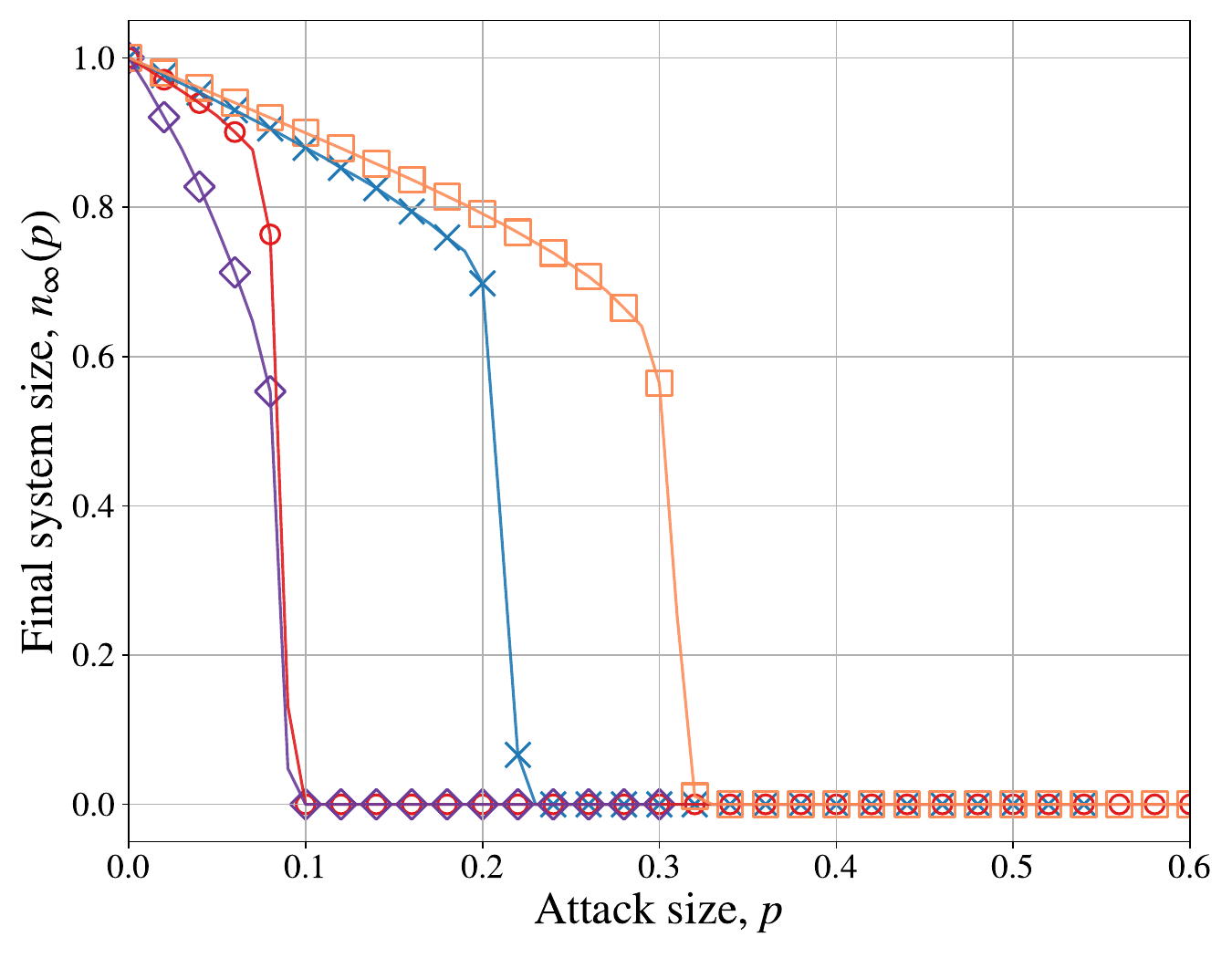}}%
\hfill%
\subfloat[ER Network: Par-Uni]{%
    \includegraphics[width=0.31\textwidth]{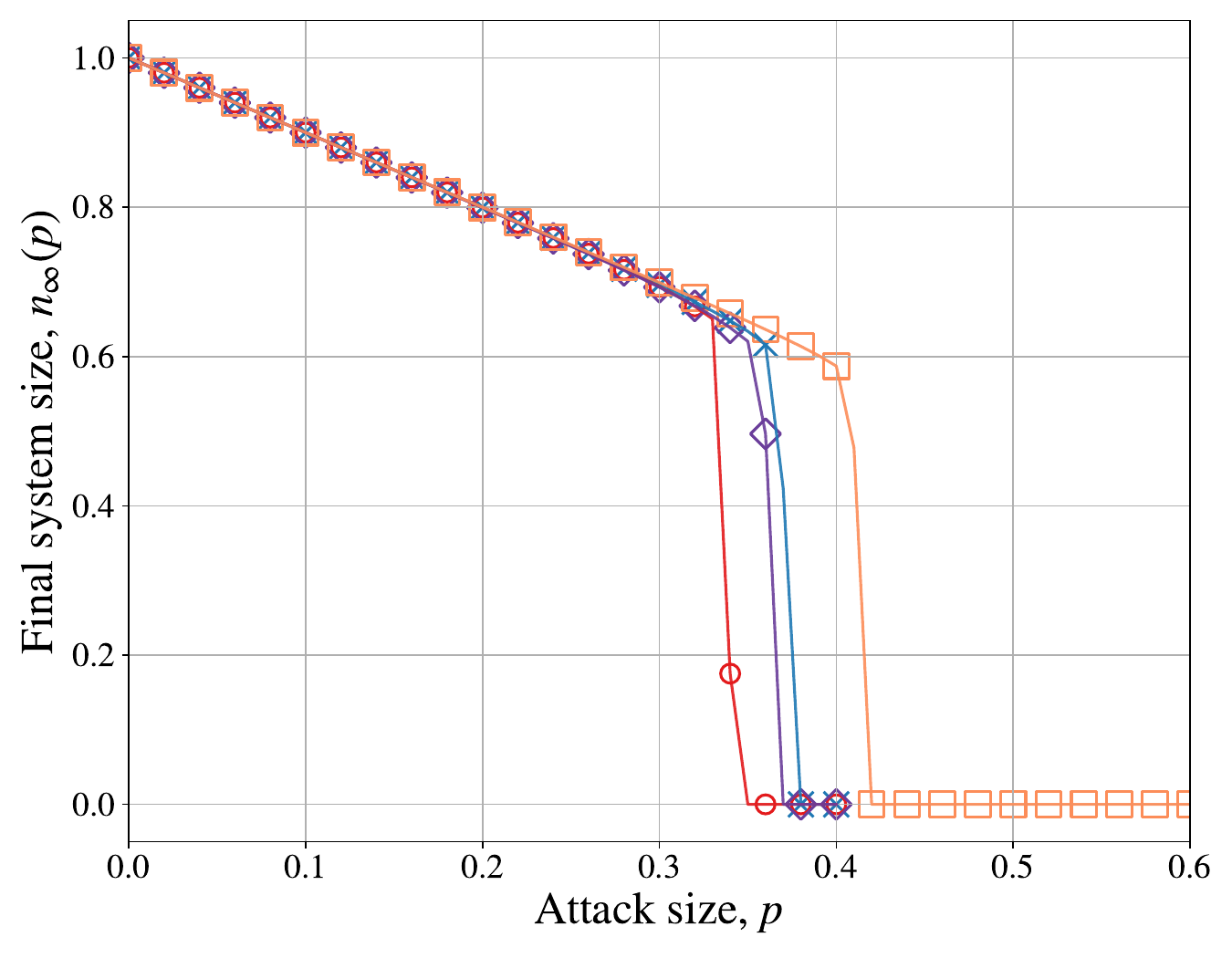}}%
\hfill%
\subfloat[ER Network: Uni-Wei]{%
    \includegraphics[width=0.31\textwidth]{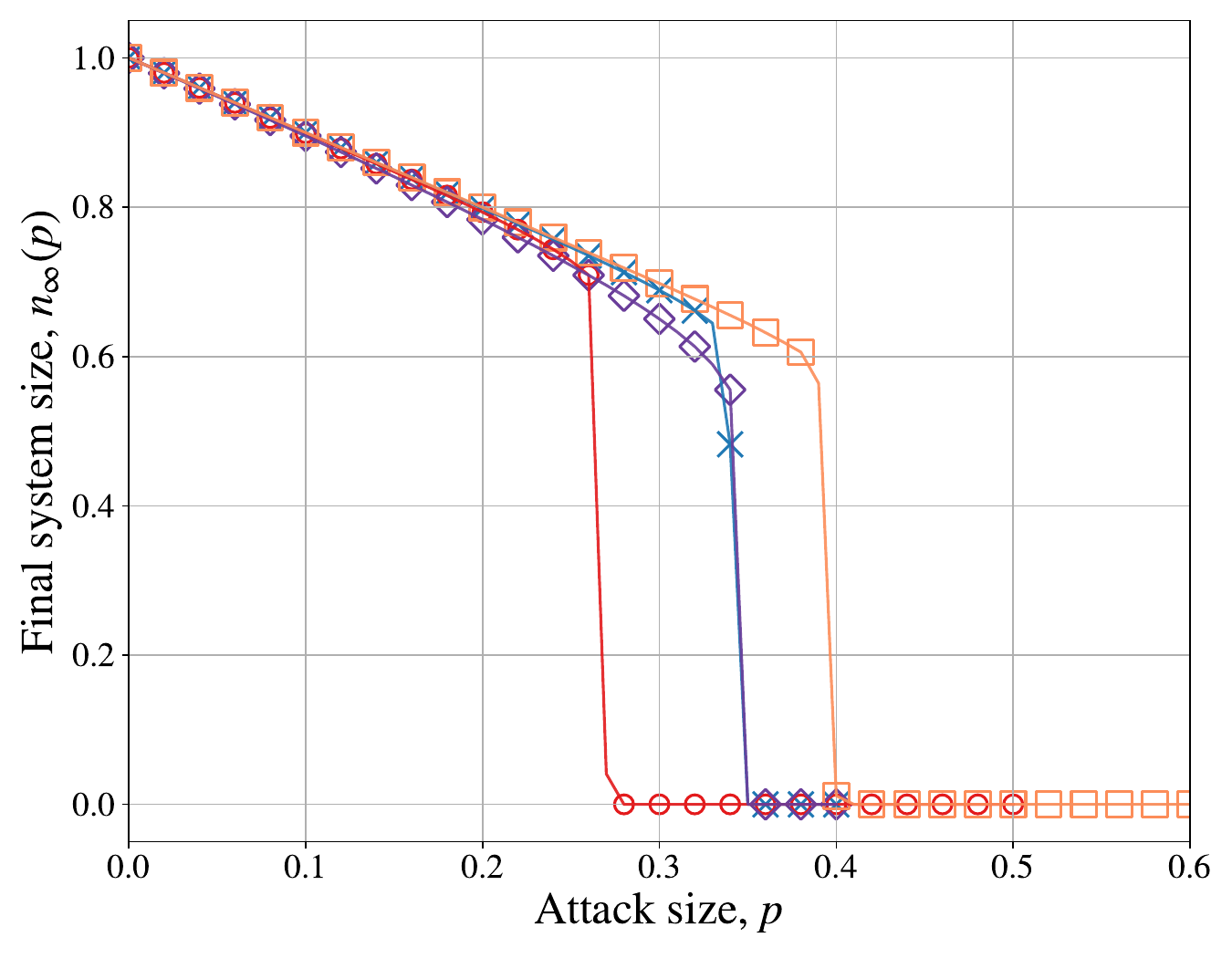}}

\vspace{0.8em}

\subfloat[SF Network: Wei-Par]{%
    \includegraphics[width=0.31\textwidth]{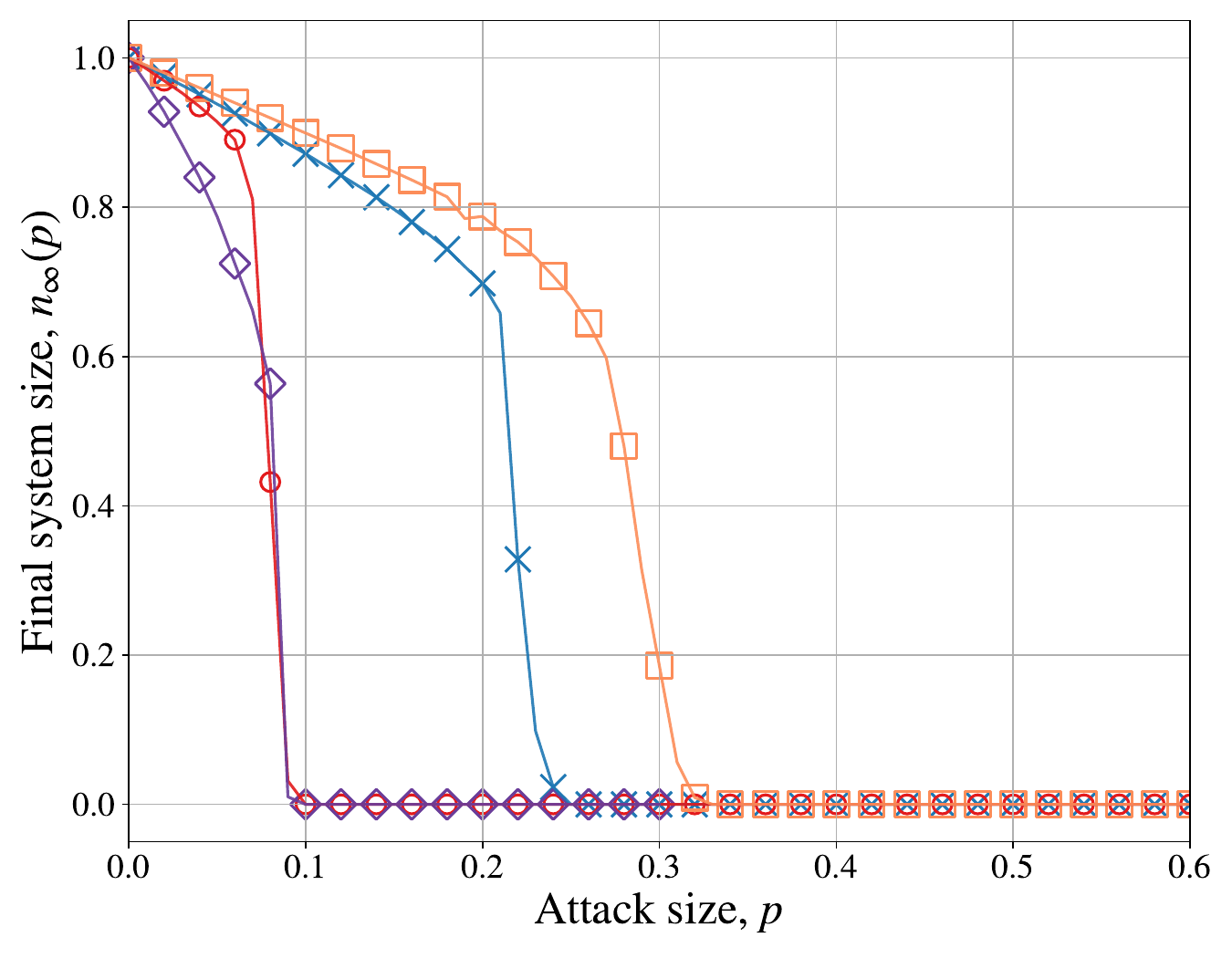}}%
\hfill%
\subfloat[SF Network: Par-Uni]{%
    \includegraphics[width=0.31\textwidth]{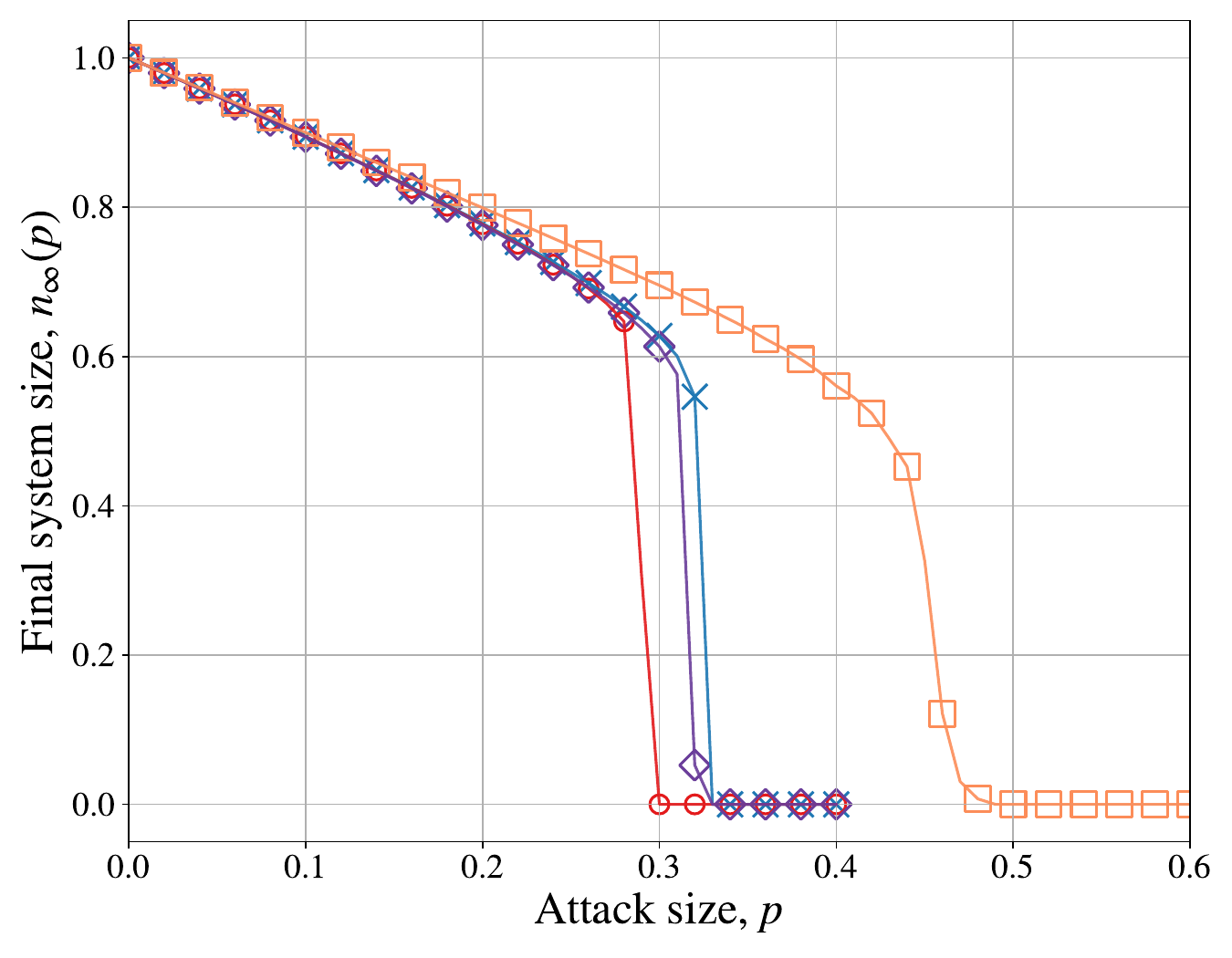}}%
\hfill%
\subfloat[SF Network: Uni-Wei]{%
    \includegraphics[width=0.31\textwidth]{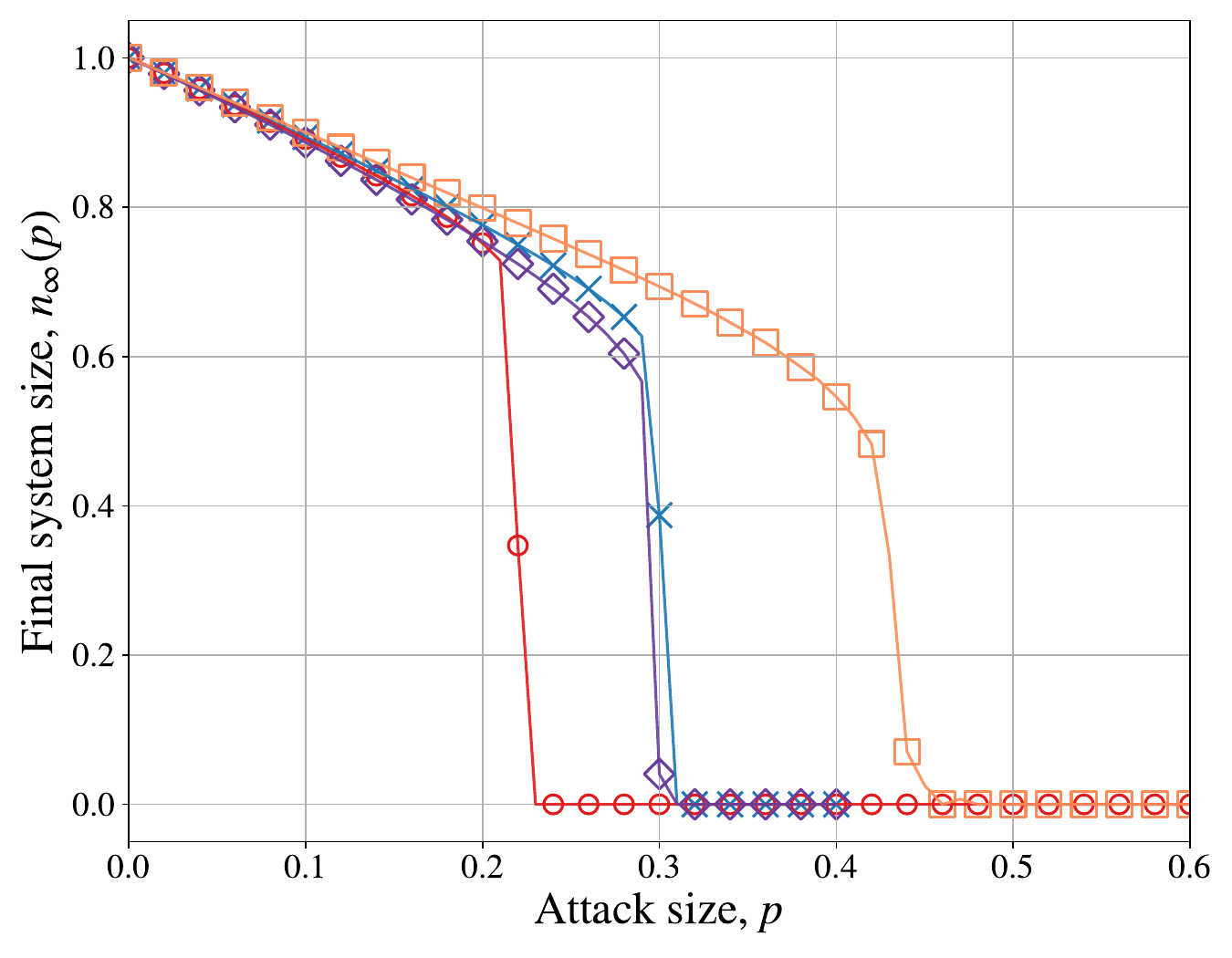}}

\vspace{0.8em}

\includegraphics[width=0.75\textwidth,trim={0cm 3cm 0cm 3cm},clip]{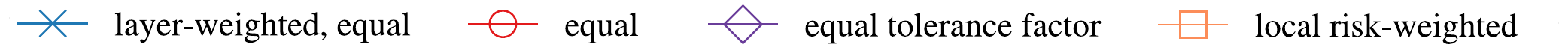}

\caption{\label{fig:local_redistribution_simulations}
Final system size for nodes surviving in both layers, $n_{AB}$, under different free-space allocation strategies with local redistribution.
The same three configurations and allocation strategies from Fig.~\ref{fig:optimal_strategy} are evaluated here, with the addition of local-risk-weighted FSA.
Panels (a)--(c) correspond to Erd\H{o}s--R\'enyi networks with mean degree 5:
(a) Weibull--Pareto, (b) Pareto--Uniform, and (c) Uniform--Weibull.
Panels (d)--(f) correspond to scale-free networks with power-law exponent $\gamma=2.55$:
(d) Weibull--Pareto, (e) Pareto--Uniform, and (f) Uniform--Weibull.
Since analytical results are not available for the local case, all points are averages over 50 simulation runs with $N=10^5$; lines connect the averaged data for visualization.
}
\end{figure*}

Figure~\ref{fig:optimal_strategy} compares the final fractions of nodes that remain functional in layer-$A$ (red triangles), in layer-$B$ (purple circles), and in both layers (blue crosses) under three free-space allocation strategies.
For both \textit{Layer-weighted equal FSA} (Figure~\ref{fig:optimal_strategy}a) and \textit{Equal FSA} (Figure~\ref{fig:optimal_strategy}b), the surviving fractions follow the line $1-p$ over a wide range of attack sizes and then drop suddenly to zero at a critical threshold.
This behavior is consistent across the initial load distributions we tested.
When each node in a layer receives the same free space, the allocation effectively fixes a common excess load threshold per node.
As we increase the attack size, we are effectively increasing the excess load per node in each layer for the initial round of flow redistribution.
As a result, until this excess load exceeds the threshold, no secondary failures occur and once it is exceeded, all nodes fail simultaneously.
The difference between the two strategies is that the \textit{Layer-weighted equal FSA} accounts for differences between layers (through expected loads and cross-layer influence).
Therefore, the collapse threshold for both layers losing functionality is aligned at the same attack size, as in Figure~\ref{fig:optimal_strategy}a.
In contrast, \textit{Equal FSA} ignores these differences and can lead to one layer failing earlier than the other, as in Figure~\ref{fig:optimal_strategy}b.
As a result, if dual-layer survival is the priority, the \textit{Layer-weighted equal FSA} provides the largest critical attack size for nodes functioning in both layers, $p^{\ast}_{AB}$, across our experiments.
If, instead, one functionality is more important, priority can be shifted toward that layer by allocating a larger share of the total free space to it (while still distributing free space equally among its nodes), which can increase the single-layer critical attack size $p^{\ast}_{A}$ or $p^{\ast}_{B}$ at the expense of the other layer and of $p^{\ast}_{AB}$.
For example, relative to Figure~\ref{fig:optimal_strategy}a ($S_A=584$, $S_B=136$), the \textit{Equal FSA} in Figure~\ref{fig:optimal_strategy}b ($S_A=S_B=360$) increases $p^{\ast}_{B}$, while reducing $p^{\ast}_{A}$ and $p^{\ast}_{AB}$.
Finally, the \textit{Equal tolerance factor} strategy in Figure~\ref{fig:optimal_strategy}c leads to earlier degradation and smaller critical attack sizes, and it consistently performs worse than the other two strategies across the load-distribution combinations we tested.

\subsection{Local Redistribution and Network Topology}
\label{sec:numerical_strategy_local}

While Sections~\ref{sec:numerical_validation}--\ref{sec:cascade_outcome_regimes} focused on the \textit{global redistribution} rule, where the load of a failed node is shared evenly among all surviving nodes in the same layer, overload effects are often more localized in practice.
In such settings, redistribution is better represented by an explicit network topology that encodes which nodes can directly share load.
For example, in the supply-chain setting discussed earlier, excess demand at a facility is more likely to be absorbed by \textit{nearby} facilities that can serve in close proximity, which can naturally be modeled through a network of local interactions.

In this subsection, we evaluate the effectiveness of different free-space allocation strategies under the \textit{local redistribution} rule.
Under the local redistribution rule, when a node fails in a given layer, its load in that layer is redistributed evenly among its surviving neighbors.
If all nodes in a connected component fail (i.e., redistribution to {\em neighboring} nodes becomes impossible), the total load of that component is redistributed evenly among all surviving nodes.
This implementation detail preserves total load across realizations and avoids artificially overestimating robustness due to lost load in isolated failed components.

The previous results in Section~\ref{sec:numerical_strategy_global} suggest that \textit{layer-weighted equal FSA} performs well under global redistribution because it reduces secondary failures: once the initial attack is applied, the system typically remains functional until a critical threshold is reached, after which it collapses completely. 
Under local redistribution, however, such secondary failures will be shaped by the network topology, since failed load is passed to immediate neighbors. 
This motivates an allocation rule that follows the same principle of limiting secondary failures, but now uses local information about potential exposure from neighboring nodes. 
Accordingly, we introduce below a fourth strategy that we refer to as \textit{local-risk-weighted FSA} (LR-FSA).

\begin{itemize}
    \item[\textbf{iv)}] \textbf{Local-risk-weighted FSA (LR-FSA):}
    For each node $i$, we estimate the overload \textit{risk}, i.e., the amount of load it would receive from its neighbors upon their failure. We approximate this by the total load it would inherit if all of its neighbors were to fail. Specifically, we define
    \[
        r_{i,A} = \sum_{v \in \mathcal{N}_A(i)} \frac{L_{v,A}}{\deg_A(v)}, 
        \qquad
        r_{i,B} = \sum_{v \in \mathcal{N}_B(i)} \frac{L_{v,B}}{\deg_B(v)},
    \]
    where $\mathcal{N}_A(i)$ (respectively, $\mathcal{N}_B(i)$) denotes the set of {\em neighbors} of node $i$ in layer-$A$ (respectively, in layer-$B$),  and $\deg_A(v) = |\mathcal{N}_A(v)|$ (respectively, $\deg_B(v)= |\mathcal{N}_B(v)|$) denotes the {\em degree} (i.e., total number of neighbors) of node $v$ in layer-$A$ (respectively, in layer-$B$).
    Put differently, $r_{i,A}$ and $r_{i,B}$ represent the  load of type-$A$ and type-$B$ that node $i$ would receive under local redistribution if all of its neighbors were to fail. To incorporate cross-layer influence, we define the effective risks
    \[
        r^{\mathrm{eff}}_{i,A} = r_{i,A} + \beta_B r_{i,B},
        \qquad
        r^{\mathrm{eff}}_{i,B} = r_{i,B} + \beta_A r_{i,A}.
    \]
    Given a fixed total free-space budget corresponding to an average free space $S_{\mathrm{tot}}$ per node, we allocate free space across node-layer pairs in proportion to these effective risks. 
    Therefore the nodes with higher load exposure or higher risk would receive higher free spaces.
\end{itemize}

We consider three initial load configurations that we also used in Section~\ref{sec:numerical_strategy_global}: Weibull--Pareto with $L_A \sim Wei(10,84.25,0.4)$ and $L_B \sim Par(5,2)$, Pareto--Uniform with $L_A \sim Par(100,5)$ and $L_B \sim U(150,200)$, and Uniform--Weibull with $L_A \sim U(80,100)$ and $L_B \sim Wei(10,225.68,2)$. 
Across all configurations, we fix $\mathbb{E}[L_A]+\mathbb{E}[L_B]=300$, $\mathbb{E}[S_A]+\mathbb{E}[S_B]=720$, and $\beta_A=\beta_B=0.2$. 
We evaluate two network families: Erd\H{o}s--R\'enyi (ER) networks with mean degree $5$, and scale-free (SF) networks with power-law degree distribution with exponent $\gamma \approx 2.55$~\cite{barabasiNS}. 
In each experiment, the two layers are generated independently so they are distinct but they share the same structural parameters within each network family.

Figure~\ref{fig:local_redistribution_simulations} reports the resulting final system size under local redistribution, focusing on the fraction of nodes that remain functional in \emph{both} layers, $n_{AB}$. 
Across all load configurations and both network families, \textit{local-risk-weighted FSA} yields the largest critical attack size $p^{\ast}_{AB}$ and the highest overall final system size. 
In several cases, the improvement in $p^{\ast}_{AB}$ relative to the second-best strategy reaches about $15\%$.
Among the remaining strategies, \textit{layer-weighted equal FSA} typically performs the second best, while its advantage relative to the \textit{equal tolerance factor} strategy is often small. 
These results suggest that when local information is available and node-level capacity assignment is feasible, LR-FSA provides a clear advantage under local redistribution. 
When such information is unavailable, \textit{layer-weighted equal FSA} remains a strong baseline. 
The observed performance gaps are generally larger in SF networks (Fig.~\ref{fig:local_redistribution_simulations}d--f), which is consistent with their higher degree heterogeneity: local exposure varies substantially across nodes, and exploiting this variation improves robustness more significantly.

\section{Conclusion}
\label{sec:discussions and conc}
In this paper, we studied overload-based cascading failures in multiplex flow networks with \textit{partial functionality}.
In this setting, the two layers share node-level resources, so the load in one layer affects the capacity in the other through the cross-layer influence factors $\beta_A$ and $\beta_B$, but a node that fails in one layer may remain functional in the other.
Under global redistribution, we derived mean-field recursive equations for the surviving fractions and excess loads, and validated them against Monte Carlo simulations for several load and free-space configurations.

The results show that partial functionality changes the structure of cascade outcomes.
Instead of a single survival state, the system can end in both-layer survival, single-layer survival, or complete collapse.
This leads to distinct critical thresholds and to phase diagrams with regimes that are absent in joint-functionality models.
The effect of cross-layer influence is also not uniform: it depends on the load and free-space distributions in the two layers.
As a result, stronger cross-layer influence can shift critical thresholds, create asymmetric layer collapses, and in some cases expand the region of dual-layer survival relative to the joint-functionality case.

We also examined free-space allocation strategies under global and local redistribution.
Under global redistribution, layer-weighted equal FSA provided the highest dual-layer robustness among the tested strategies.
Under local redistribution on Erd\H{o}s--R\'enyi and scale-free networks, the proposed local-risk-weighted FSA achieved the best performance in the tested configurations.

Several extensions remain open.
First, the local redistribution case could be studied more directly for specific network topologies.
This would help clarify how degree heterogeneity, neighborhood structure, and layer-to-layer correlations affect the cascade dynamics under partial functionality.
Second, the free-space allocation problem could be treated in a more systematic way.
If an exact characterization of the optimal allocation is not feasible, one could instead derive upper and lower bounds on the best attainable robustness and use these bounds to estimate the optimality gap of different FSA strategies.
Lastly, additional attack strategies, such as highest-load attacks, could be considered to develop more effective free-space allocation strategies and to analyze how cascade outcomes change under targeted attacks.

\begin{acknowledgments}

This work was supported in part by the Air Force Office of Scientific Research (AFOSR) Grant \# FA9550-22-1-0233.
O. İrsoy gratefully acknowledges the support of Knight Fellowship through the IDeaS Center at Carnegie Mellon University for the 2024-2025 academic year.
O. İrsoy gratefully acknowledges the support of  David Barakat and LaVerne Owen-Barakat Fellowship through the College of Engineering at Carnegie Mellon University for the 2025-2026 academic year.
\end{acknowledgments}

\nocite{REVTEX42Control}
\nocite{apsrev42Control}

\bibliographystyle{apsrev4-2}
\bibliography{ref}

\end{document}